# A solution of the Boltzmann transport equations for spin and charge transport in a solid. Spin Proximity effect.

*V. Zayets*[*]

*Spintronic Research Center, National Institute of Advanced Industrial Science and Technology (AIST), Umezono 1-1-1, Tsukuba, Ibaraki, Japan. E-mail: v.zayets@aist.go.jp*

Abstract.

*A solution of the modified Boltzmann transport equations is found, which describes features of the spin and charge transport in a solid. Two modifications of the Boltzmann transport equation were introduced. The first modification describes the fact that a delocalized electrons can either be of the running-wave type or the standing-wave type and electrons of different types contribute differently to the transport. The second modification includes the fact that the direction of the electron spin may not be conserved after frequent electron scatterings. The origins and features of the spin proximity, spin injection and spin detection effects are described. An enhancement of spin detection and spin injection efficiencies in the vicinity of an interface are predicted. The physical mechanism of an enlargement of spin accumulation due to the conventional Hall effect is described.*

## 1. Introduction

The classical model [1-3] of the electron gas in a solid assumes that the spins of all electrons in the electron gas are aligned and all the electrons occupy two bands (Fig.1(a)-(c)). The electron spins of one band are aligned in the opposite direction to the spin direction of the electrons of the other band. This classical model is called the model of spin-up/spin-down bands. According to this model, in the case when there is no spin accumulation in the electron gas (Fig. 1(a)), the number of electrons with the spin directed "up" is equal to the number of electrons with the spin directed down. The actual "up" and "down" directions are not specified and any two opposite directions could be chosen as the "spin-up" and "spin-down" directions. The reason of this ambiguity is the assumption of the model of spin-up/spin-down bands that the electron gas without spin accumulation is fully described by projections of electron spins and actual spin directions of electrons are not relevant in this case. When a magnetic field is applied, the spins of all electrons are realigned (Fig. 1(b)). The spin direction of each electron becomes well-defined and it is either parallel or antiparallel to the direction of the magnetic field. The Fermi energy of "spin-up" and "spin-down" bands in the magnetic field is still the same. Because of the difference in the Zeeman energy for electrons with spin direction along and opposite to the magnetic field, the bottoms of "spin-up" and "spin-down" bands are at different energies and the number of electrons the "spin-up" and "spin-down" bands is different (See Fig. 1(b)). Therefore, there is a spin accumulation in the electron gas when a magnetic field is applied. This effect is called the Pauli paramagnetism [1]. There is other case when the electron gas may be spin-polarized. Spin injection causes a spin accumulation in the electron gas. Spin injection may occur, for example, due to an illumination of the metal by circular-polarized light or due to a flow of a spin polarized current from another metal. The model of spin-up/spin-down bands assumes that even a negligibly-small spin injection causes realignment of spins of all electrons in the electron gas. Spins of all electrons are realigned to be either parallel or antiparallel to the spin direction of injected electrons (Fig.3(c)). Due to the spin injection, the number of electrons, whose spin is parallel to spin of the injected electrons, becomes larger and the energy of the Fermi level of electrons of this band becomes higher. There is a spin accumulation in this case, but it is of a different kind than the spin accumulation induced by a magnetic field (Compare Fig. 1(b) and Fig. 1(c)).

In [4] it has been proved that the direction of the electron spin may not be conserved after a spin-independent scattering. This fact has to be included in a model of the electron gas in a solid and the model of spin-up/spin-down bands should be modified. Because of the spin rotations after frequent scatterings, spins of electrons can not be aligned into two opposite directions. Figures 1(d)-(f) show distributions of spin directions in an electron gas, when the fact of frequent spin rotations is not ignored. The electrons in the electron gas can be divided into two groups. In one group there is no spin alignment. In this group electron spins are distributed with an equal probability in all directions. In an electron gas without spin accumulation all electrons are in this group (Fig.1 (d)). In the other group all electron spins are aligned in one direction. Electrons can be in this group only in the case when there is a spin accumulation in an electron gas. A spin accumulation induced by a magnetic field or a spin accumulation due to a spin injection are not distinguishable and they are shown in Fig. 1(e) and (f).

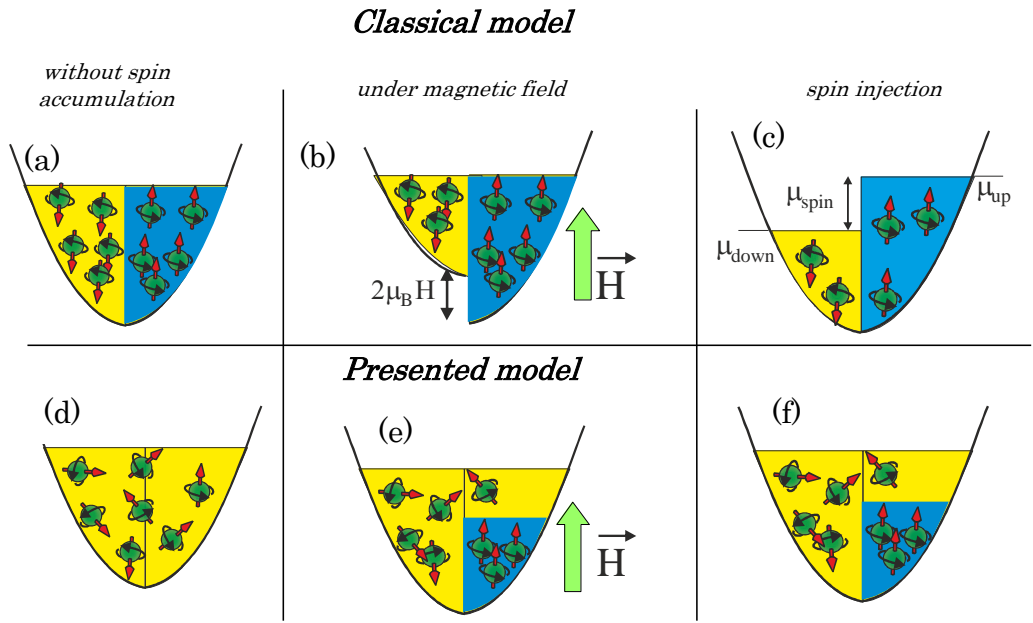

Fig.1. *Distribution of spin directions in an electron gas (a),(b),(c) for the model of spin-up/spin-down bands and (d),(e),(f) for the presented model. (a),(d) case when there is no spin accumulation. (b),(e) case when there is a magnetic field (Pauli paramagnetism). (c),(f) case when there is a spin injection (accumulation)*

Comparing spin projections for the spin distributions according to a model of spin-up/spin-down bands (Fig. 1(a)-(c)) and the proposed model (Fig. 1(d)-(f)), the similarities of both models are noticeable. However, there are significant differences between the models in the description of the charge and spin transport. According to the classical model, the spin-up and spin-down bands are almost independent and they can be described by individual chemical potentials [5,6]. Only seldom spin-flip scatterings cause the exchange of electrons between the bands. The model of spin-up/spin-down bands assumes that the transport of electrons in each band is independent [5] and the electrons of different bands may have different conductivities [6]. However, because spin may be rotated after a scattering, even if at some moment the electrons of different spin directions would have different Fermi energies, in time of a few picoseconds (Chapter 8 of [4]) the frequent scatterings will mix up all electrons. Therefore, in an electron gas with or without spin accumulation, one Fermi energy and one chemical potential should describe all electrons independently on their spin direction.

In this paper the Boltzmann transport equations are solved including the fact that the spin direction may be rotated after spin-independent scatterings. The features of the spin and charge transport such as the spin injection, spin detection, spin proximity, the Hall effect, the Spin Hall effect are calculated and analyzed.

In Chapter 2 the spin and charge transport equations are derived. Solving these equations the spin injection and the Spin Proximity effect are studied in Chapter 3 and the spin detection is studied in the chapter 4.

In an equilibrium the electron gas is spin-polarized in a ferromagnetic metal and it is not spin-polarized in a non-magnetic metal [1-6]. At a contact between a ferromagnetic and a non-magnetic metals, some spin accumulation from the ferromagnetic metal diffuses into the non-magnetic metal. The spin polarization in the ferromagnetic metal near the contact becomes smaller and the spin polarization in the non-magnetic metal becomes non-zero. This effect is called the Spin Proximity effect. When a drift current flows from a ferromagnetic to non-magnetic metal, the spin accumulation from the ferromagnetic metal may be drifted into the non-magnetic metal. This effect is called the spin injection In chapter 3 the origin and features of the Spin Proximity and spin injection effects are described.

The Spin Proximity effect should be distinguished from the Magnetic Proximity effect [7,8]. The Magnetic Proximity effect describes the change of magnetic properties of the localized d- or f- electrons in the vicinity of an interface, because of the interlayer exchange coupling. In contrast, the Spin Proximity effect describes the change of the spin polarization of the electron gas of the delocolized sp- electrons, because of the spin diffusion through the interface. Because of the sp-d exchange interaction between the localized and delocalized electrons, both effects are correlated.

Still they can be distinguished, because of their different effective lengths. The effective length of the Magnetic Proximity effect and the interlayer exchange coupling does not exceed a few monolayers [7-9]. In contrast, the effective length of the Spin Proximity effect is longer than several spin-diffusion lengths [9-11]. Because of the different effective lengths, the Spin Proximity effect and the Magnetic Proximity effect can be distinguished experimentally [9]. It should be noticed that the model of spin-up/spin-down bands is unable to explain the Spin Proximity effect.

The origin and features of the spin detection are described in the chapter 4. A spin accumulation diffuses from regions of a higher spin accumulation to regions of a lower spin accumulation. A diffusion spin current consists of a spin-polarized current and a current, which is not spin-polarized. These two currents diffuse in opposite directions. In the bulk, the two currents are of the same magnitude and there is only a spin diffusion without any charge diffusion. However, for example near a contact, these two currents become unequal and the spin current causes a charge accumulation along direction of the spin diffusion. The charge accumulation is proportional to the diffusion spin current and it can be measured by measuring the voltage along two diffusion points. As consequence, the corresponded spin accumulation can be measured by measuring the voltage [12,13]. This effect is called the spin detection.

All delocolized electrons in the electron gas can be divided into two groups: the running-wave electrons and the standing-wave electrons. Running-wave electrons move. Since they move, they transport their own charge and spin in the direction of their movement. In contrast, the standing-wave electrons do not move. They can transport the spin and the charge only due to scattering from one state to another. Correspondingly, two kinds of electron currents can be distinguished. The first type of current is the current of the running-wave electrons, which describes the transport of the spin and charge during the movement of the running-wave electrons between scatterings. The second current is the scattering current, which describes the transfer of the spin and charge because of scatterings. Both the running-wave and standing-wave electrons contribute to the scattering current. In Chapter 5 the general features of the scattering and running-wave-electron currents are discussed.

In the Chapter 6 possible modifications of the Boltzmann equations for spin and charge transport are discussed. Several facts are included into the equations. The first fact is that electrons can be either of the standing-wave type or the running-wave type. The electrons of different types contribute differently to the conductivity. The second fact is that there are two major mechanisms of the electron conductivities: the running-wave conductivity and the scattering conductivity. The third fact is that the electron gas can be spin-polarized and the electrons can be divided into two groups according to their spin directions (Fig. 1 (d)-(f)).

In the Chapter 7 the Boltzmann transport equations are solved for the running-wave electron current for the case of transport in the bulk of a metal with a low density of defect. It is shown that the detection conductivity is zero and there is no spin detection effect in this case. It is also demonstrated that the contributions the electrons with energies below and above the Fermi energy to the spin injection conductivity are of opposite sign. It leads to the fact the spin injection efficiency is small in a metal and it is determined by a gradient of the density of the states at the Fermi energy in the metal. The spin injection efficiency is substantial in the bulk of the semiconductor and it is of opposite sign for p- and n- semiconductors.

In Chapter 8 the Boltzmann transport equations are solved for a scattering current. It is shown that the scattering current is substantial only in case when there is a difference in electron scattering probabilities between two opposite directions. For example, in the vicinity of contact between two metals, the scattering current may be large even if it is negligibly small in the bulk of each metal. It is because the scattering probabilities towards the contact and in the opposite direction are different in the vicinity of the contact. Another example of the scattering current is the Spin Hall effect. The electrons, which are scattered into the left side and into the right side with respect to the flow direction of the drift current, may experience opposite direction of the effective magnetic field of the spin-orbit interaction. This may cause a different electron scattering probability to the left and right directions. The difference depends on the spin direction. This difference may cause a flow of the spin-polarized scattering current perpendicularly to the drift current. This effect is called the Spin Hall effect [14,15].

In Chapter 9 the important role of the mean-free path $\lambda_{mean}$ for the spin and charge transport is described. The influence of $\lambda_{mean}$ on transport becomes substantial in two cases. The first case is when the mean-free path $\lambda_{mean}$ becomes comparable to the average distance between defects. The second case is when an electron distance from an interface is comparable or less than $\lambda_{mean}$. In both cases the number of running-wave electrons decreases and the number of standing-wave electrons increases. As a consequence, the conductivity reduces and importantly the spin properties of the conductivity are significantly modified. The detection conductivity becomes non-zero and the injection conductivity becomes larger.

In Chapter 10 the spin and charge transport in the vicinity of a contact is described. The transport in the vicinity of a contact is significantly different than that in the bulk. If in the bulk of a metal the major transport mechanism is the running-wave electron current, near a contact the scattering current may dominate. The detection and injection conductivities are larger in the vicinity of a contact.

In Chapter 11, a unique feature of the conventional Hall effect to enlarge the spin polarization in an electron gas is described. The Hall effect occurs when a magnetic field is applied perpendicularly to the direction of a drift current. In this case both the electrons and the holes turn out from the flow direction and they are accumulated at a side edge of the wire. When the electron gas is spin-polarized, the spin is also accumulated. The charge of electrons and holes is opposite. Therefore, in a drift current the electrons and holes diffuse in opposite directions. As a consequence, in a magnetic field the electrons and holes are turned into the same direction and they both are accumulated at the same side of the sample. In a metal the numbers of electrons and holes participating in the transport is about the same and about the same amounts of electrons and holes are accumulated at a side of the sample. In total the charge of the accumulated electrons is canceled by the opposite charge of accumulated hole. The total accumulated charge in a metal due to the Hall effect is small and the Hall voltage is small. In contrast, the spin direction of the electrons and the hole in an electron gas is the same and the Hall effect causes a substantial spin accumulation at one edge of a metal wire.

## 2. Transport equations

The delocalized electrons of the electron gas occupy quantum states. Each state is distinguished by the direction of its wavevector in the Brillouin zone, its energy E and its spatial symmetry. Due to the Pauli excursion principle, each state can be occupied maximum by two electrons of opposite spin. In [4] we define a state as an "empty", "spin" or "full" state, when the state is filled by no electrons or one electron or two electrons of opposite spin, respectively. The spin of a "full" and an "empty" states is zero and the spin of a "spin" state is ½. The energy distributions of "full", "empty" and "spin" states are described by the spin statistic (Chapter 4 of Ref. [4]). The states of higher energy (at least 5kT above the Fermi energy) are not occupied by electrons and almost all of them are "empty" states with no spin. Almost all states of lower energy (at least 5kT below the Fermi energy) are occupied by two electrons and almost all of them are the "full" states with no spin. A state, the energy of which is near the Fermi energy, may be filled by one electron and have spin ½.

All delocalized electrons of the electron gas may be divided into two groups, which are named the Time-Inverse Symmetrical (TIS) and Time-Inverse Symmetrical TIA assemblies. It has been shown [4] that the number of electrons in each assembly is conserved after a spin-independent scattering. The TIA assembly contains only "spin" states with the same spin direction. The TIS assembly contains "full" states, "empty" states and "spin" states. The "spin" states of the TIS assembly have all spin directions with equal probability (Fig. 1(d), (e), (f)).

The spin and charge distributions in samples of different geometries can be calculated from the spin and charge transport equations. It is a set of two differential equations with two variables: the chemical potential $\mu$ and the spin polarization *sp*.

The transport equations can be derived from the continuity equations for spin and charge. The continuity equations describe the conservation laws for spin and charge, which require that the amount of spin and charge at each special point may change only when electrons are converted between assemblies or when electrons defuse from a point to a point.

Electrons can be converted from the TIA to the TIS assembly because of spin relaxation. Electrons can be converted back from the TIS to TIA assembly because of spin pumping. For example, the spin pumping occurs when a magnetic field applied to the electron gas [4].

The spin polarization *sp* defines the number of the "spin" states $n_{spin,TIA}, n_{spin,TIS}$ in the TIA and TIS assemblies from the total number of the "spin" states $n_{spin}$ in the electron gas as [4]

$$n_{spin,TIA} = sp \cdot n_{spin} \quad n_{spin,TIS} = (1-sp) \cdot n_{spin} \tag{1}$$

The total number of the "spin" states in both assemblies $n_{spin}$ can be calculated from the spin statistics [4] and the density of states of a metal. For example, in a metal, for which density of states is a constant in the vicinity of the Fermi energy, $n_{spin}$ is can be calculated as

$$n_{spin} = D \cdot kT \cdot (1.14463 + 0.26055 \cdot sp) \tag{2}$$

where D is the density of states at the Fermi energy. It should be noticed that $n_{spin}$ only slightly depends on the spin polarization *sp*. It decreases about 20 % as the spin polarization increases from 0 to 100%. In contrast, $n_{spin}$ may significantly depend on a charge accumulation. For example, in the case of a non-degenerate n-semiconductor $n_{spin}$ increases exponentially with linear increase of the Fermi energy. In a metal it could be assumed that $n_{spin}$ only weakly depends on the charge accumulation.

Due to spin relaxation and spin pumping, the number of the spin states in each assembly changes as [4]

$$\frac{\partial n_{TIA}}{\partial t} = -\frac{\partial n_{TIS}}{\partial t} = -\frac{n_{TIA}}{\tau_{spin}} + \frac{n_{TIS}}{\tau_{pump}} \tag{3}$$

where $\tau_{spin}$ is the spin relaxation time and $\tau_{pump}$ is the spin pumping effective time.

Substituting Eq. (1) into Eq. (3) gives

$$\frac{\partial n_{TIA}}{\partial t} = n_{spin} \left( \frac{1-sp}{\tau_{pump}} - \frac{sp}{\tau_{spin}} \right) \tag{4}$$

In an equilibrium the number of the spin states in the assemblies does not change. Therefore, the spin polarization $sp_0$ of the equilibrium can be calculated from the condition

$$\frac{\partial n_{TIA}}{\partial t} = n_{spin} \left( \frac{1-sp_0}{\tau_{pump}} - \frac{sp_0}{\tau_{spin}} \right) = 0 \tag{5}$$

which gives

$$\tau_{pump} = \tau_{spin} \frac{1-sp_0}{sp_0} \tag{6}$$

Substituting Eq. (6) into Eq. (4) gives the conversion rate between assemblies as

$$\frac{\partial n_{TIA}}{\partial t} = -\frac{\partial n_{TIS}}{\partial t} = \frac{n_{spin}}{\tau_{spin}} \left( sp_0 \frac{1-sp}{1-sp_0} - sp \right) \tag{7}$$

Next, we calculate the spin and charge currents, which flows along the gradients of the chemical potential $\mu$ and the spin polarization *sp*. As was mentioned above, the model of spin-up/spin-down bands incorrectly assumes that there are two independent chemical potentials for spin-up and spin-down electron bands, there are two independent energy distributions for spin-up and spin-down electrons and there could be two independent currents for spin-up and spin-down electrons [5,6]. It is an incorrect assumption because frequent electron scatterings quickly mix up all electrons of all possible spin polarizations ensuring existing of only one energy distribution for all electrons of different spin distributions. For this reason the electrons of the TIA and TIS assemblies (spin-polarized and spin-unpolarized electrons) always have the same Fermi energy and the same chemical potential $\mu$.

Even having the same chemical potential, the electrons of the TIA and TIS assemblies diffuses independently along the gradients of the chemical potential $\mu$ and the spin polarization *sp*. The currents of electrons of the TIA and TIS assemblies can be expressed as

$$\vec{J}_{TIA} = \sigma_{\mu,TIA} \nabla\mu + \sigma_{sp,TIA} \cdot kT \cdot \nabla sp$$
$$\vec{J}_{TIS} = \sigma_{\mu,TIS} \nabla\mu + \sigma_{sp,TIS} \cdot kT \cdot \nabla sp \tag{8}$$

where conductivities $\sigma_{\mu,TIA}$ $\sigma_{sp,TIA}$ $\sigma_{\mu,TIS}$ $\sigma_{sp,TIS}$ can be calculated by solving the Boltzmann transport equations (See Chapters 6-9). The currents flowing along $\nabla\mu$ are drift currents and the currents flowing along $\nabla sp$ are diffusion currents [6].

The electrons of the TIA assembly are spin-polarized. When they defuse they transport spin and charge. In contrast, the electrons of the TIS assembly are not spin-polarized and they only transport the charge. Therefore, the spin and charge currents can be calculated as

$$\vec{J}_{charge} = \vec{J}_{TIA} + \vec{J}_{TIS} = \sigma_{charge} \nabla\mu + \sigma_{detection} \cdot sp \cdot kT \cdot \nabla sp$$
$$\vec{J}_{spin} = \vec{J}_{TIA} = \sigma_{injection} sp \cdot \nabla\mu + \sigma_{spin} \cdot kT \cdot \nabla sp \tag{9}$$

where charge, spin, detection and injection conductivities are defined as

$$\sigma_{charge} = \sigma_{\mu,TIS} + \sigma_{\mu,TIA} \quad \sigma_{detection} = (\sigma_{sp,TIS} + \sigma_{sp,TIA})/sp$$
$$\sigma_{injection} = \sigma_{\mu,TIA}/sp \quad\quad \sigma_{spin} = \sigma_{sp,TIA} \tag{10}$$

The charge conductivity $\sigma_{charge}$ is the conventional conductivity of the metal, which describes a charge current flow along a gradient of the chemical potential (the Ohm's law). The spin-diffusion conductivity $\sigma_{spin}$ describes the conventional spin diffusion along a gradient of the spin polarization. The spin polarization of an electron current is not necessarily equal to the spin polarization of an electron gas (See Chapter 7 and Fig. 7). The injection conductivity $\sigma_{injection}$ defines a ratio of how much the spin polarization of the drift current smaller than the spin polarization of the electron gas. It describes the ability of a drift current to transport a spin accumulation. Usually $\sigma_{injection}$ is small in a metal and it is large in a semiconductor (See Chapter 7). The detection conductivity $\sigma_{detection}$ describes the fact that a charge may accumulate along a spin diffusion. This electrical charge can be measured and the magnitude of a spin current can be evaluated. Usually $\sigma_{detection}$ is small in the bulk of a semiconductor and a metal (See Chapter 7). It become substantial in the vicinity of an interface (See Chapter 11). It should be noted that the multiplier *sp* is introduced in front of $\sigma_{detection}$ and $\sigma_{injection}$ based on the requirement of the same time-inverse symmetry of both sides of Eqs. (10). The conductivities $\sigma_{charge}$, $\sigma_{spin}$ $\sigma_{injection}$ and $\sigma_{detection}$ generally depend on the spin polarization *sp* of the electron gas and a charge accumulation. In the Chapters 7-9 the conductivities $\sigma_{charge}$, $\sigma_{spin}$ $\sigma_{injection}$ and $\sigma_{detection}$ are calculated by solving the Boltzmann transport equation.

The continuity equations for electrons of the TIA and TIS assemblies are given as

$$\nabla \cdot \vec{J}_{TIA} = -\frac{\partial n_{TIA}}{\partial t}$$
$$\nabla \cdot \vec{J}_{TIS} = -\frac{\partial n_{TIS}}{\partial t} = \frac{\partial n_{TIA}}{\partial t} \tag{11}$$

Substituting Eqs. (9) into Eqs (12), the *Spin/Charge Transport Equations* are obtained as

$$\nabla \cdot \left( \sigma_{charge} \cdot \nabla \mu + \sigma_{detection} \cdot kT \cdot sp \cdot \nabla sp \right) = 0$$

$$\nabla \cdot \left( \sigma_{injection} \cdot sp \cdot \nabla \mu + \sigma_{spin} \cdot kT \cdot \nabla sp \right) = \frac{n_{spin}}{\tau_{spin}} \left( sp - \frac{1-sp}{1-sp_0} sp_0 \right) \quad (12)$$

The transport equations (12) are a set of non-linear differential equations, which in general should be solved numerically. However, in some cases it is possible to solve it analytically. In a simple case of a spin diffusion in the bulk of a non-magnetic metal, in which $\nabla \mu = 0$, $\sigma_{detection} = 0$, $sp_0 = 0$ the transport equations (12) are simplified to

$$\nabla \cdot \left( \sigma_{spin} \cdot kT \cdot \nabla sp \right) = sp \frac{n_{spin}}{\tau_{spin}} \quad (13)$$

In the case when $\sigma_{spin}$ does not depend on *sp*, Eq. (13) can be simplified to the Helmholtz equation

$$\nabla^2 sp = \frac{sp}{\lambda_{spin}^2} \quad (14)$$

where the spin diffusion length is calculated from Eq. (13) as

$$\lambda_{spin} = \sqrt{\frac{\sigma_{spin} \cdot kT \cdot \tau_{spin}}{n_{spin}}} \quad (15)$$

### 3. Spin Proximity effect. Spin injection.

When two metals, which have different spin polarization, are in contact, in the vicinity of the contact interface some spin-polarized electrons (electrons of TIA assembly) diffuse from the metal of a larger spin accumulation into the metal of a smaller spin polarization. For example, in the bulk of a non-magnetic metal the electron gas is not spin-polarized and it is spin-polarized in the bulk of a ferromagnetic metal. Near contact between ferromagnetic and non-magnetic metals the electron gas becomes spin-polarized in the non-magnetic material, because of the spin diffusion into this region from the neighboring region of the ferromagnetic metal. Correspondingly, the spin polarization in the ferromagnetic metal becomes smaller near the contact than in the bulk, because spins were diffused out of there. This effect is called the Spin Proximity effect.

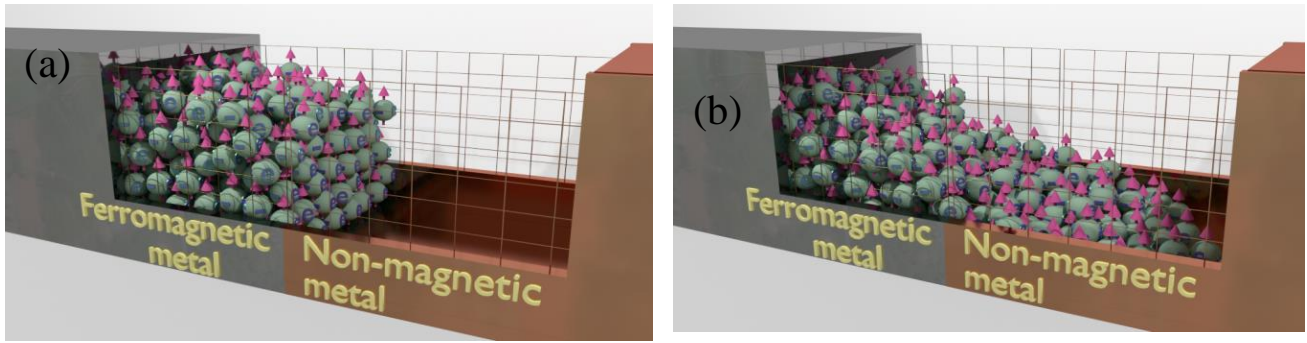

*Fig. 2 The spin accumulation at a contact between a non-magnetic and a ferromagnetic metals. (a) The model of spin-up/down bands. It assumes that all spin-polarized electrons are contained only inside the ferromagnetic metal. It is as if an invisible wall prevents the spin accumulation from diffusion into the non-magnetic metal. (b) The presented model. Spin accumulation diffuses from the ferromagnetic metal into the non-magnetic metal.*

When a drift current flows through a contact between ferromagnetic and non-magnetic metals, an additional spin accumulation is drifted from the ferromagnetic metal into the non-magnetic metal. For the opposite direction of the current, the spin accumulation in the non-magnetic metal is drifted back into the ferromagnetic metal. This effect is called the spin injection. It is important to emphasize that the spin injection just changes the amount of spin accumulation in the non-magnetic metal, which already was there when there was no current.

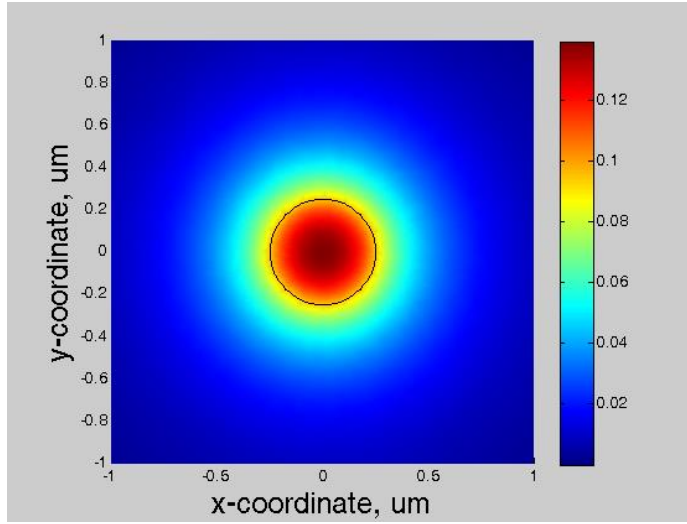

*Fig. 3. Calculated spin polarization of an electron gas in a ferromagnetic cylinder embedded in a non-magnetic metal. The diameter of the cylinder is 0.5 μm. The boundary of the cylinder is shown by the black line. There is a significant spin diffusion from the ferromagnetic metal into the non-magnetic metal.*

There is a significant amount of experimental evidences on the existence of the Spin Proximity effect [8-11] and the spin injection effect [16-19]. The model of spin-up/spin-down bands is unable to explain the Spin Proximity effect. Within the model of spin-up/spin-down bands it is assumed that in a contact between a ferromagnetic metal and a non-magnetic metal a spin accumulation is contained only inside the ferromagnetic metal and it does not diffuse into the non-magnetic metal (Fig.2a). Only when a voltage is applied to the contact and a drift current flows through the contact, some of the spin accumulation may be drifted into the non-magnetic metal. This story contradicts with experimental observations [10, 11]. The presented model explains the spin injection and the spin transport through the contact differently. Even without a current nothing prevents a diffusion of a spin accumulation from the ferromagnetic metal into the non-magnetic metal. The spin accumulation decays from the ferromagnetic metal through the contact deep into the non-magnetic metal (Fig. 2b). The drift current only modifies this distribution.

The following examples demonstrate the Spin Proximity effect and spin injection for a contact between a ferromagnetic and a non-magnetic metals. The spin polarization of the ferromagnetic metal $sp_0$ equals to 0.6. Except for the spin polarization, all other parameters of the metals are the same. The conductivity $\sigma_{charge}$ is 2E7 S/m. The density of the states at the Fermi level is 2E22 1/cm3/eV. The spin life time is 30 ps. The spin-diffusion conductivity $\sigma_{spin} = 1.3\ \sigma_{charge}$. The injection conductivity $\sigma_{injection} = 0.7\ \sigma_{charge}$. The detection conductivity $\sigma_{detection} = 0$ (See Chapter 7). Temperature is the room temperature. Data were calculated using the Finite Difference method. It was assumed that the conductivity near the interface is the same as the bulk conductivity and there is no contact conductivity (See Chapter 9).

Fig.3 shows the calculated distribution of spin polarization of an electron gas in a ferromagnetic cylinder embedded in a non-magnetic metal. The diameter of the cylinder is 0.5 μm. The spin polarization is largest at the center of the cylinder and it becomes smaller at the edge. Even at the center of the ferromagnetic cylinder the largest spin polarization is about 0.13, which is significantly reduced from the bulk spin polarization of 0.6. Around the cylinder the electron gas in the non-magnetic metal is significantly spin-polarized. Therefore, the Spin Proximity effect causes the significant reduction of spin polarization of the electron gas in the ferromagnetic cylinder and it causes a spin polarization of the electron gas in the non-magnetic metal.

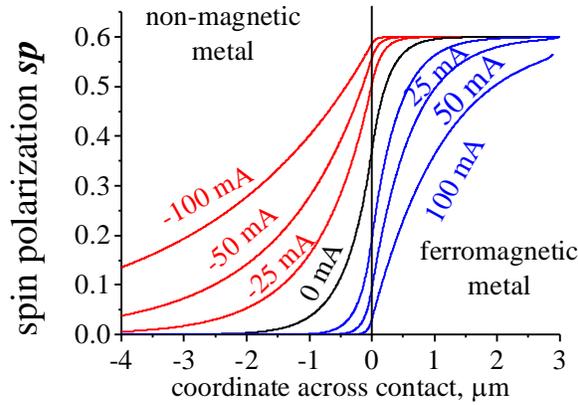

*Fig.4. Spin polarization of the electron gas across the contact between a ferromagnetic and a non-magnetic metals for different values of charge current flowing through the contact. Even without the current (black line) there is a spin accumulation in the non-magnetic metal. For a negative current (red lines) the amount of the spin accumulation becomes larger. For a positive current (blue lines) the amount of the spin accumulation in the non-magnetic metal becomes smaller.*

Figure 4 shows the calculated distribution of the spin polarization across the contact between the ferromagnetic and the non-magnetic metals for different values of the drift current flowing across the contact. The contact area is 0.01 μm². The constants of the metal are the same as were used for the calculations of Fig. 3. In all cases the spin polarization changes smoothly (without a step) from one metal to other metal. Even without a drift current, the electron gas is spin-polarized inside the non-magnetic metal and the spin polarization in the ferromagnetic metal is smaller than its bulk spin polarization. This is because of the Spin Proximity effect. For a negative drift current (red lines) the spin accumulation is drifted from the ferromagnetic metal into the non-magnetic metal. The spin polarization in the non-magnetic metal becomes larger and the spin polarization in the ferromagnetic metal becomes near 0.6, which is its bulk spin polarization. For a positive drift current (blue lines) the spin accumulation from the non-magnetic metal is drifted back into the ferromagnetic metal and most of the spin accumulation is depleted. In the ferromagnetic metal the spin accumulation is drifted deep inside of the bulk and near the contact interface the spin accumulation is depleted.

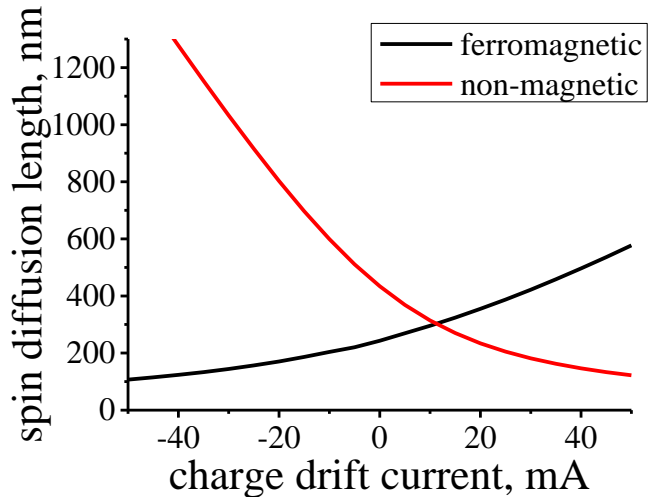

*Fig.5 Effective spin length as a function of the charge current flowing through a contact between a ferromagnetic and non-magnetic metals.*

The distribution of the spin polarization in each metal can be fitted as

$$sp_{nonMag}(x) = A_n \cdot e^{\frac{x}{\lambda_{spin,nonMag}}}$$

$$sp_{ferro}(x) = sp_0 - A_f \cdot e^{-\frac{x}{\lambda_{spin,ferro}}} \quad (16)$$

where x is the coordinate across the contact, $sp_0$ is the bulk spin polarization of the ferromagnetic metals, $A_n$ and $A_f$ are constants and $\lambda_{spin,nonMag}$, $\lambda_{spin,ferro}$ are the effective spin diffusion lengths in the non-magnetic and ferromagnetic metals, respectively.

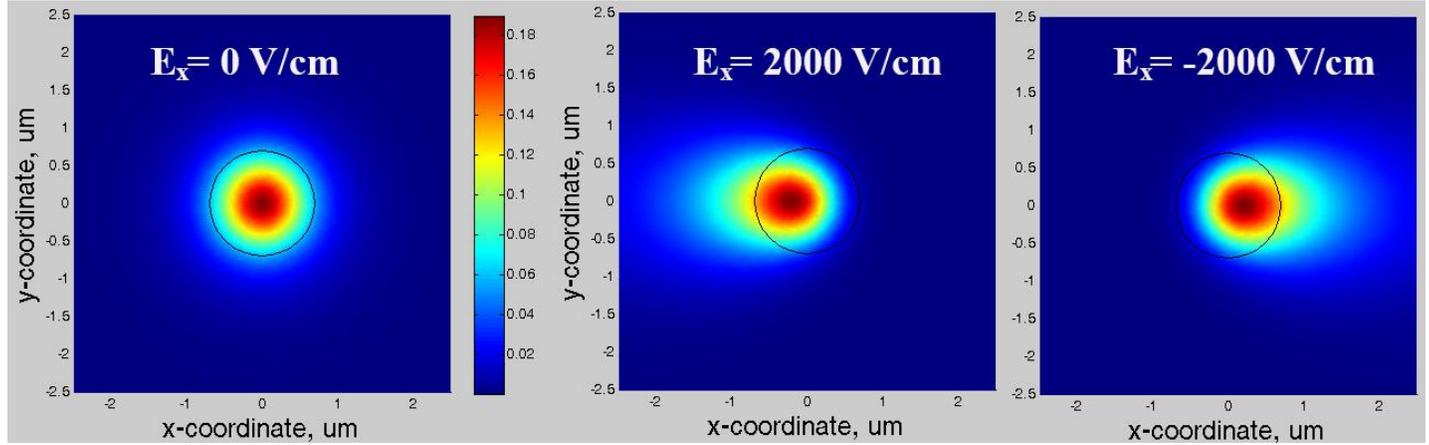

*Fig.6. The drift of the spin polarization in a metal under an applied electrical field. The spin polarization of metal is induced by circularly polarized light. The light beam is Gaussian.*

Figure 5 shows the effective spin diffusion length in the non-magnetic and ferromagnetic metals as the function of the drift current. When the spin accumulation is injected from the ferromagnetic to the non-magnetic metal (negative drift current), the effective spin length increases in the non-magnetic metal and it decreases in the ferromagnetic metal. This effect is called the spin gain/damping [6]. It occurs because of the conversion of the spin drift current into the diffusion spin current along the flow of the drift current. When the flow of the converted spin current is in the same direction as the flow the diffusion current, the converted spin current assists the flow of the diffusion spin current and the spin diffusion length increases. When the direction is the opposite, the spin diffusion current is damped and the spin diffusion length decreases. The decrease and increase of the spin diffusion length depending of the direction of the spin current is well-confirmed and measured experimentally [10,11].

The spin injection may occur not only between two metals, but within the same metal as well. Figure 6 shows the calculated spin polarization induced in the non-magnetic metal by circularly-polarized light at different voltages applied in the plane of the metal. The constants of the metal are the same as were used for the non-magnetic metal calculated in Figs. 3- 5 except $\sigma_{injection} = 0.9 \cdot \sigma_{charge}$ When a voltage is not applied in-plane of the metal, the distribution of the spin polarization is Gaussian. When the voltage is applied, the spin polarization is drifted along the voltage. The drift only occurs in metals in which $\sigma_{injection} \neq 0$. The drift of the spin polarization similar to that calculated in Fig.6 was observed experimentally in n-GaAs (Fig.4 of Ref. [10]).

The spin polarization of a spin current in a semiconductor is almost equal to the spin polarization of the electron gas. In contrast, in a metal the spin polarization of a spin current is significantly smaller than the spin polarization of the electron gas. This fact will be proved in Chapter 7 from a solution of the Boltzmann transport equations. A simplified explanation of this fact can be given as follows. In an n-type semiconductor the charge and spin are transported by electrons. The negatively-charged electrons move from a "-"source towards a "+" drain and the spin is drifted in the same direction. In a p-type semiconductors the charge and spin are transported by holes. The positively-charged holes move from a "+" towards a "-" drifting the spin in the same direction. Therefore, the spin drifting direction is opposite in the n-type and p-type semiconductors. This is the reason of the opposite signs of $\sigma_{injection}$ for n- and p-semiconductors. The electrical conduction in a metal can be seen as a drift of electrons and hole in opposite directions (Fig.7). Since the charge of electrons and holes is opposite, the electron and hole currents transport the charge in the same direction, but they transport the spin in opposite directions. Since the number of electrons and holes is about the same in a metal, the spin drifting by electrons and holes nearly compensate each other. This causes a small spin-polarization of a drift current in a metal and a small value of $\sigma_{injection}$. The sign of $\sigma_{injection}$ in a metal may be either negative or positive depending on whether there are more electrons or holes in the metal. The later condition is determined by the sign of the gradient of the density of the states of the metal at the Fermi energy. The sign of $\sigma_{injection}$ is important for an effective spin injection. For example, for an effective spin injection from a ferromagnetic metal to a non-magnetic metal, the sign of $\sigma_{injection}$ should be the same for the ferromagnetic and non-magnetic metals. The polarity of the applied voltage should correspond to the sign of $\sigma_{injection}$. The sign $\sigma_{injection}$ can be evaluated from the sign of the Hall voltage in the metal (See Chapter 12).

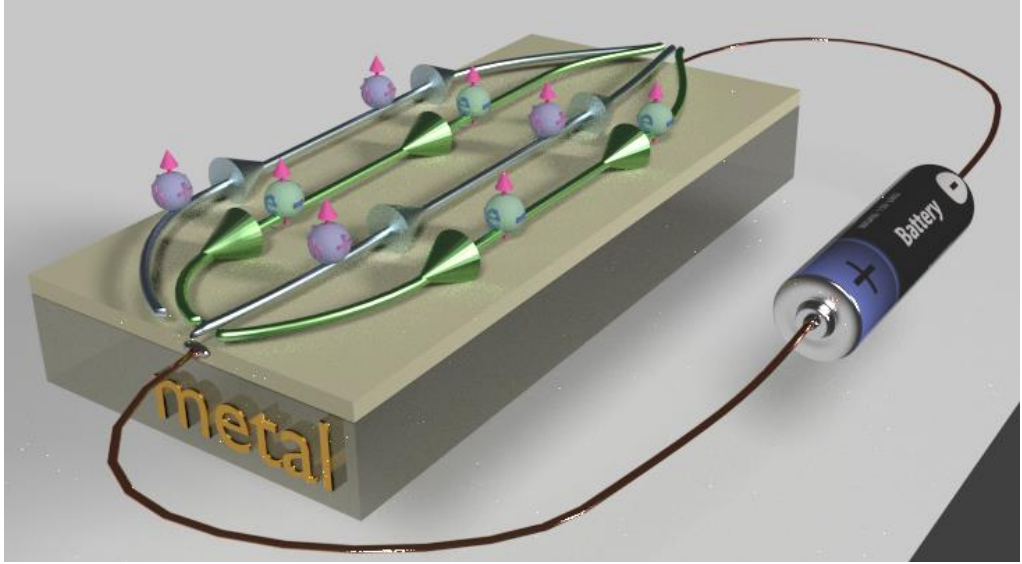

*Fig.7. A drift current in a metal. There are two kinds of carriers: negatively- charged electrons (green balls) and positively-charged holes (blue balls). Both carriers may transport spin. If the direction of the charge transport is the same for both carriers, the direction of spin transport is opposite. This is the reason why the spin polarization of the drift current in a metal is significantly smaller than the spin polarization of an electron gas.*

An efficient spin injection is important for a variety of different application. For example, the largest spin-transfer torque is one of main required parameters for a cell of magnetic random-access memory (MRAM). The largest spin-transfer torque is only possible in the case when there is an efficient spin injection between the electrodes of the MRAM cell [4]. It is possible only when metals with the largest value of $\sigma_{injection}$ are used as electrodes of the MRAM cell.

It should be noticed that all above descriptions of the spin injection and the Spin Proximity effects were well-matched to experimental observations. For example, in Ref. [11] the spin injection from Fe into n-GaAs was studied and the spin accumulation in the GaAs was directly imaged using the Hanle effect. When the direction of the drift current is such that the electrons flow from the Fe into the n-GaAs, an increase of the spin accumulation in the n-GaAs is observed (Fig. 7(c), Fig.7 (e) of Ref. [11]) and the spin diffusion length is elongated in the n-GaAs. For the opposite direction of the current (Fig. 7(d), Fig.7 (f) of Ref.[11]), the spin accumulation in the n-GaAs decreases and the spin diffusion length is shortened.

## 4. Spin Detection

A spin accumulation in an electron gas can be detected by measuring the magnetic field induced by the accumulated spins. However, this magnetic field is rather small and practically it is difficult to measure such a small magnetic field. The spin detection is an effect, which allows to measure a spin current and a spin accumulation electrically by measuring an electrical voltage [12,13]. The spin detection voltage is induced by a charge, which may be accumulated along the diffusion direction of a spin current. The following describes the reason why the charge is accumulated along the spin diffusion and what is the origin of the spin detection effect.

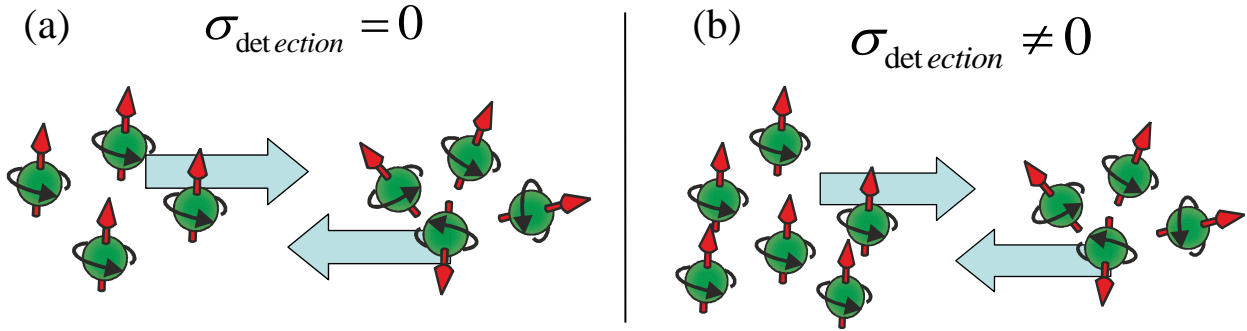

*Fig.8. A diffusion spin current in metal. The spin-polarized electrons (electrons of the TIA assembly) diffuse from the region of a higher spin accumulation (on left side) to the region of a lower spin accumulation (on right side). The spin-unpolarized electrons (electrons of the TIS assembly) diffuse in the opposite direction from right to left. (a) The case when equal amounts of spin-polarized electrons and spin-unpolarized electrons diffuse in opposite directions. There is no charge accumulation. (b) The case when the number spin-polarized electrons, which diffuse from left to right, is larger than the number of spin-unpolarized electrons, which diffuse in the opposite direction. This causes an increase of the number of electrons at the left side and a decrease of the number of electrons at the right side. This means the diffusion spin current causes a charge accumulation at the left side and a charge depletion at the right side of the metal.*

In the electron gas the charge and spin are transported by electrons. An electron has spin and charge and it transports both the charge and spin simultaneously. However, a current of electrons, which transports only spin but not the charge, is possible. It is the case when the spin polarized electrons (electron of the TIA assembly) diffuse in one direction and exactly the same amount of the electrons, which are not spin-polarized (electron of the TIS assembly), diffuse in the opposite direction. When the spin-polarized and spin-unpolarized currents are the same, in total there is no charge diffusion, but there is a spin diffusion. Such a case is only possible in a material in which the conductivities of the electrons of the TIA and TIS assemblies are the same. For example, it is the case of conduction in the bulk of a high-purity conductor. In a conductor with defects or in the vicinity of an interface there is a slight difference of conductivities of the TIA and TIS assemblies (See Chapters 10,11). In this case the spin diffusion is accompanied by a small charge diffusion. In contrast to a spin current, which needs only a source, a charge current needs both a source and a drain. A charge current without a drain builds up a charge accumulation, which induces a strong electrical field. The induced electrical field causes a charge current in the direction opposite to the direction of the diffusion charge current. The build up of the charge accumulation is stopped when the drift charge current becomes equal to the diffusion charge current and in total there is no charge current. By measuring the voltage induced by the build-up charge accumulation, the amount of spin current and the spin accumulation can be evaluated.

Figure 8 explains the origin of the spin detection. The figure shows a diffusion spin current flowing from region of higher spin polarization (on left) to the region of greater spin polarization (on right). Figure 8(a) shows the spin current in a metal, in which the conductivities of electrons of the TIA and TIS assemblies are equal. In this case, equal amounts of spin-polarized electrons and spin-unpolarized electrons move in the opposite directions and there is no diffusion charge current. For example, 4 spin-polarized electrons move left and 4 spin-unpolarized electrons moved right. There is a spin movement from left to right, but there is no charge movement. Figure 8(b) shows the spin current in a metal where the conductivity of electrons of the TIA assembly is larger than the conductivity of electrons of the TIS assembly. In this case, two more electrons flow from left to right, there is a diffusion charge current and there is a charge accumulation at right side.

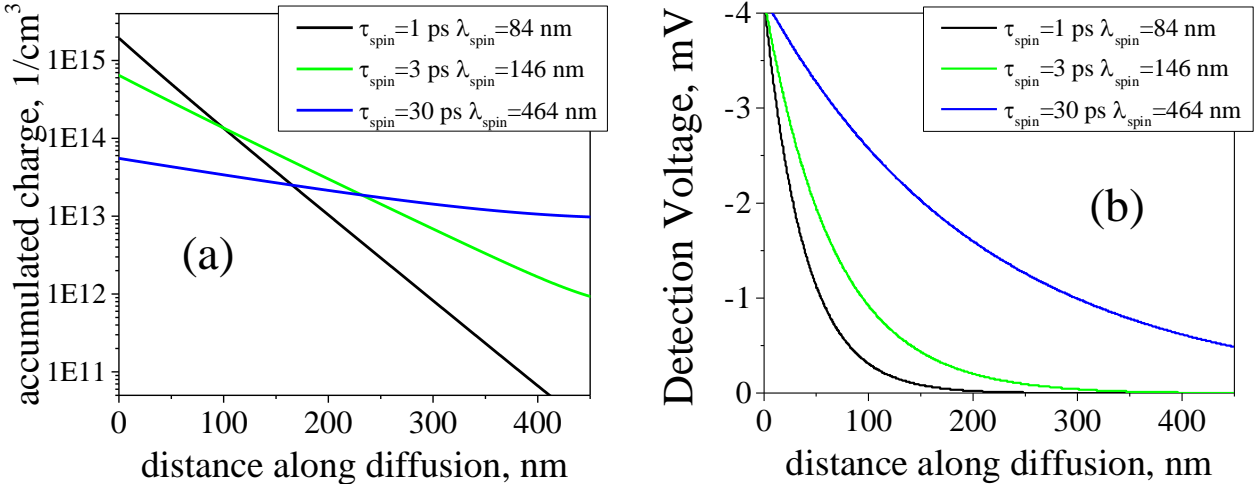

*Fig.9. (a) accumulated charge and (b) detection voltage along a spin diffusion in a metal wire. Metals with different spin life time $\tau_{spin}$ and spin diffusion length $\lambda_{spin}$ are shown. At zero distance the spin polarization **sp** equals 0.6 in all cases.*

The sensitivity of the spin detection is linearly proportional to the value of $\sigma_{detection}$. As was mentioned above the value of $\sigma_{detection}$ is negligibly small in the bulk of conductors, but it becomes significant in the vicinity of a contact between two different conductors (See Chapter 11).

Figure 9 shows the accumulate charge and the detection voltage along the spin diffusion in a non-magnetic wire. The conductivity of the wire $\sigma_{charge}$ is 2E7 S/m, the detection conductivity $\sigma_{detection} = 0.9 \cdot \sigma_{charge}$ the injection conductivity $\sigma_{injection} = 0$. The spin-diffusion conductivity $\sigma_{spin} = 1.15 \cdot \sigma_{charge}$. The density of states at the Fermi energy is 2E22 1/cm^3/eV. The cases of metals with different spin life times and spin diffusion lengths are shown. The diffusion starts at spin polarization **sp**=0.6, which corresponds in graphs to the zero diffusion distance. In metals with shorter spin life time the charge accumulation is larger at zero diffusion distance. Approximately one order shorter spin life time corresponds to one order larger charge accumulation. In contrast, the detection voltage does not depend on the spin life time. As can be seen from Fig. 9(b) the voltage between the point of the zero diffusion distance and a distant point is the same for metals of different spin life time. It should be noticed that in this case the detection voltage does not depend also on the drift current. Therefore, the detection voltage can be reliably used to detect the magnitude of a spin accumulation.

In the bulk of a conductor with a low density of defect the detection conductivity equals zero $\sigma_{detection} = 0$ and there is no spin detection effect (See Chapter 7). In a conductor with a larger number of defects the detection conductivity becomes non-zero, but it is still small (See Chapter 9). In the vicinity of a contact between two conductors or an interface between a conductor and dielectric, the detection conductivity may become substantial (See Chapter 10).

## *5. Running-wave electrons and standing-wave electrons. Current of running-wave electrons and scattering current.*

The wave nature of electrons significantly influences the spin and charge transport in a solid. Since an electron is a wave, it is not necessary that an electron with non-zero speed moves and transports a spin and charge. A standing wave has a non-zero speed, but it does not move. For a calculation of the transport it is important to distinguish two classes of the electrons: the standing-wave electrons and the running-wave electrons. A standing-wave electron can be imagined as an electron bouncing back and forward between two defects. Also, a standing wave electron can be viewed as two electrons, which move in opposite directions. In order to build up a standing-wave electron two defects are not necessarily required, but one defect is often sufficient. Also, a standing-wave electron can form at an interface. The probability a standing-wave electron can be formed at a defect or an interface can be characterized by a parameter called the reflectivity R. The reflectivity R is defined as the ratio of the probability for an electron to be a standing-wave type to the probability to be a running wave type.

In the bulk of a high-crystal-quality metal with a low density of defects and dislocations, most of the electrons are running-wave electrons. In a metal with a larger amount of defects or in the vicinity of an interface, there is a substantial amount of the standing-wave electrons. In the vicinity of a contact between two metals, there are both running-wave electrons and standing-wave electrons, because the reflectivity of such a contact is usually small. In contrast, in the vicinity of higher resistance contacts like a tunnel barrier contact or a semiconductor-metal contact almost all electrons are standing-wave electrons.

There are two major mechanisms of electron transport in an electron gas. A running-wave electron transports the spin and charge, because it moves. If there is a smaller amount of electrons moving in one direction than the amount of electrons moving in the opposite direction, there is a flow of charge and/or spin in this direction. Such a type of current is defined as the running-wave-electron current. This transport describes the transport of the charge and the spin by electrons between scatterings. It is important to emphasis that only the running-wave electrons, but not standing-wave electrons contribute to this current.

In the electron gas the electrons are continuously scattered from one quantum states to another quantum state. When the probability for an electron to be scattered in one direction is higher than the probability to be scattered in the opposite direction, there is a flow of the spin and charge in this direction. This current is defined as the scattering current. Both the standing-wave and running-wave electrons may contribute to this current. Usually the scattering current is significantly less efficient than the running-wave-electron current for the charge and spin transport.

## 6. Boltzmann transport equations

The goal of studying the Boltzmann Transport Equations in this manuscript is to calculate the charge, spin-diffusion, detection and injection conductivities, which are used in the Spin/Charge Transport Equations (Eqs.(12)). The Boltzmann equations describe a temporal evolution of the distribution function $F(\vec{r},\vec{p},t)$. The distribution function $F(\vec{r},\vec{p},t)$ describes the probability to find an electron at a point $\vec{r}$ with pulse $\vec{p}$ at time t.

We have modified the Boltzmann Transport Equations in order to include several facts, which are essential for the description of the spin and charge transport. We consider the fact that each quantum state can be occupied either by two electrons of opposite spins, by one electron or the state can be unoccupied. The spin states can be either in the TIS assembly (spin directions are equally distributed in all directions) or in the TIA assembly (there is only one spin direction for all electrons of the assembly). Therefore, four distribution functions are required to describe the transport. The distribution functions for "spin" states of the TIA assembly, "spin" states of the TIA assembly, "full" and "empty" states $F_{spin,TIA}, F_{spin,TIS}, F_{full}, F_{empty}$ are used, which describe the probabilities for a quantum state to be filled by one, two or no electrons, correspondingly. The condition that an electron can be only in one of these four quantum states gives

$$F_{spin,TIA} + F_{spin} + F_{full} + F_{empty} = 1 \tag{17}$$

The second fact, which was used in the modified Boltzmann transport equations, is that there is a spin relaxation in the electron gas and there is a conversion between electrons of different assemblies. The third fact is that the running-wave electrons and the standing-wave electrons contribute differently to the transport. The forth fact is that there are two possible currents in an electron gas: the running-wave-electron current and the scattering current.
Including all these facts, the Boltzmann transport equation is given as

$$\frac{dF_i}{\partial t} = \left(\frac{dF_i}{\partial t}\right)_{RunningWave} + \left(\frac{dF_i}{\partial t}\right)_{Scattering} + \left(\frac{dF_i}{\partial t}\right)_{force} + \left(\frac{dF_i}{\partial t}\right)_{conversion} + \left(\frac{dF_i}{\partial t}\right)_{relaxation} \tag{18}$$

where "i" labels the distribution functions for "spin" of the TIA, "spin" of TIS, "full" and "empty" states; $\left(\frac{dF_i}{\partial t}\right)_{RunningWave}$ is the term, which describes a change of the distribution function due to the movement of the running-wave electrons; $\left(\frac{dF_i}{\partial t}\right)_{Scattering}$ is the scattering term, which describes the changing of the distribution function due to the movement of the electrons by scatterings from one quantum state to another quantum state; $\left(\frac{dF_i}{\partial t}\right)_{force}$ is the force term, which describes the change of the distribution function due an external field; the conversion term $\left(\frac{dF_i}{\partial t}\right)_{conversion}$ describes the electron conversion between assemblies, because of the spin relaxation or the spin pumping; the relaxation term $\left(\frac{dF_i}{\partial t}\right)_{relaxation}$ describes the relaxation of the distribution function to an equilibrium distribution function due to the electron scatterings.

Except for some special cases it is safe to assume that an external perturbation (applied electrical and magnetic fields, a thermo gradient, a gradient of spin accumulation, a sp-d exchange effective field and a spin-orbit effective magnetic field) is sufficiently small so that under the perturbation the distribution function is only slightly different from the distribution function in an equilibrium. In this case the distribution function can be represented as

$$F_i = F_{i,0} + F_{i,1} \qquad (19)$$

where $F_{i,0}$ is the distribution function at an equilibrium and $F_{i,1}$ describes a small deviation from the equilibrium such that for any point of the phase space the following condition is valid:

$$F_{i,0} \gg F_{i,1} \qquad (20)$$

It could be further assumed that the speed of the relaxation of the distribution function into the equilibrium is linearly proportional to a deviation of the distribution function from the equilibrium. Then, the relaxation term can be calculated as

$$\left(\frac{dF_i}{\partial t}\right)_{relaxation} = -\frac{F_i - F_{i,0}}{\tau_k} = -\frac{F_{i,1}}{\tau_k} \qquad (21)$$

where $\tau_k$ is the momentum relaxation time. Since electrons are constantly scattered between assemblies at a high rate, $\tau_k$ is the same for both assemblies and $\tau_k$ is the same for distributions of "spin", "full" and "empty" states. Usage of the relaxation term in form of Eq. (21) is called the relaxation-time approximation.

The electrons of the TIA assembly are converted into the TIS assembly, because of the different mechanisms of the spin relaxation. The rate of the conversion due to the spin relaxation is proportional to the number of spin states in the TIA assembly [1-3]:

$$\left(\frac{dF_{spin,TIS}}{\partial t}\right)_{conversion} = -\left(\frac{dF_{spin,TIA}}{\partial t}\right)_{conversion} = \frac{F_{spin,TIA}}{\tau_{spin}} \qquad (22)$$

where $\tau_{spin}$ is the spin life time.

The electrons of the TIS assembly may be converted into the TIA assembly due to the spin pumping. For example, the spin pumping occurs when a magnetic field is applied to the electron gas [4]. The conversion rate due to the spin pumping is linearly proportional to the number of spin states in the TIS assembly:

$$\left(\frac{dF_{spin,TIA}}{\partial t}\right)_{conversion} = -\left(\frac{dF_{spin,TIS}}{\partial t}\right)_{conversion} = \frac{F_{spin,TIA}}{\tau_{pump}} \qquad (23)$$

where $\tau_{pump}$ is the effective spin pump time. $\tau_{pump}$ is inversely proportional to the magnitude of the applied magnetic field [4]. It should be noticed that the integration Eqs. (22) and (23) over all states gives Eqs. (4) and (5).

Combining Eqs. (20) and (21), the conversion term is given as

$$\left(\frac{dF_{spin,TIS}}{\partial t}\right)_{conversion} = -\left(\frac{dF_{spin,TIA}}{\partial t}\right)_{conversion} = \frac{F_{spin,TIA}}{\tau_{spin}} - \frac{F_{spin,TIS}}{\tau_{pump}} \qquad (24)$$

## 7. Current of running-wave electrons

The current of running-wave electrons is the major transport mechanism in the bulk of a metal. It occurs because of the movement of electrons between scatterings. It is a very efficient transport mechanism. The standing-wave electrons do not contribute to this transport mechanism.

The solution of the Boltzmann transport equation for the running-wave electrons looks similar to the solution of the classical Boltzmann equation for the electron gas in a solid, which can be found in many textbooks on Solid State Physic. However, the classical Boltzmann equation ignores the fact that only the running-wave electrons contribute to the current of running-wave electrons and the standing-wave electrons do not contribute.

The current of running-wave electrons occurs, because of the movement of electrons in space. The movement of electron literally means that if at time t an electron is at point x, at time t+dt the electron will be at point $x + v_x \cdot dt$, where $v_x$ is the x-axis projection of the electron speed. If at a point the running-wave electrons are described by a distribution function F(x), the change of F(x) due to the movement of the running-wave electrons along the x-direction can be described as

$$dF(x) = F(x - v_x \cdot dt) - F(x) \qquad (25)$$

In the case of a short time interval dt, Eq. (25) can be simplified as

$$\frac{\partial F}{\partial t} = -v_x \cdot \frac{\partial F}{\partial x} \qquad (26)$$

Taking into the account that an electron can move not only in the x-direction, but in any direction gives the running-wave term of the Boltzmann equation as

$$\left(\frac{\partial F}{\partial t}\right)_{RunningWave} = -\vec{v} \cdot \nabla F \qquad (27)$$

The Boltzmann transport equation Eq. (16) only with the running wave term and the relaxations term in a static case is simplified to

$$\frac{dF_i}{\partial t} = \left(\frac{dF_i}{\partial t}\right)_{RunningWave} + \left(\frac{dF_i}{\partial t}\right)_{relaxation} = 0 \qquad (28)$$

Using the relaxation-time approximation Eqs. (19), (21) and substituting Eq. (27) into Eq. (28) gives

$$-\frac{F_{i,1}}{\tau_k} - \vec{v} \cdot \nabla\left(F_{i,0} + F_{i,1}\right) = 0 \qquad (29)$$

The solution of Eq. (29) using the condition (20) is

$$F_{i,1} = -\tau_k \cdot \vec{v} \cdot \nabla F_{i,0} \qquad (30)$$

From Eq. (26), the current of the running-wave electrons can be found as

$$\vec{j}_i = \frac{e}{(2\pi\hbar)^3} \iiint \vec{v} \cdot (F_{i,0} + F_{i,1}) \cdot d\vec{p} = \frac{e}{(2\pi\hbar)^3} \iiint \vec{v} \cdot F_{i,1} \cdot d\vec{p} = \frac{-\tau_k e}{(2\pi\hbar)^3} \cdot \iiint \vec{v} \cdot (\vec{v} \cdot \nabla F_{i,0}) \cdot d\vec{p} \qquad (31)$$

Integration in Eq. (31) is only over running-wave electrons. The standing wave electrons should not be included into the integration.

In the case of a transport in the bulk of an isotropic metal, Eq. (31) can be simplified. We define the angle θ as the angle between the electron movement direction and $\nabla F_{i,0}$. The number of electrons in the assembly can be calculated as

$$n_i = \frac{1}{(2\pi\hbar)^3} \iiint F_{i,0}(E) \cdot d\vec{p} = \int dE \cdot D(E) \cdot F_{i,0}(E) \cdot \int_0^\pi 0.5 \cdot \sin(\theta) \cdot d\theta = \int D(E) \cdot F_{i,0}(E) \cdot dE \qquad (32)$$

where D(E) is the density of the states.

Using Eq. (32) in Eq. (31), the currents flowing along and perpendicularly to $\nabla F_{i,0}$ is calculated as

$$j_{i,\parallel} = -\tau_k e \cdot \int D(E) \cdot dE \cdot \int_0^\pi 0.5 \cdot \sin(\theta) \cdot d\theta \cdot |\vec{v}| \cdot \cos(\theta) \cdot \left(|\vec{v}| \cdot |\nabla F_{i,0}| \cdot \cos(\theta)\right) = -\frac{e \cdot \tau_k}{3} \cdot \int D(E) \cdot |\vec{v}|^2 \cdot |\nabla F_{i,0}| dE$$

$$j_{i,\perp} = -\tau_k e \cdot \int D(E) \cdot dE \cdot \int_0^\pi 0.5 \cdot \sin(\theta) \cdot d\theta \cdot |\vec{v}| \cdot \sin(\theta) \cdot \left(|\vec{v}| \cdot |\nabla F_{i,0}| \cdot \cos(\theta)\right) = 0$$

(33)

or

$$\vec{j}_i = -\frac{e \cdot \tau_k}{3} \cdot \int D(E) \cdot |\vec{v}|^2 \cdot \nabla F_{i,0} \cdot dE$$

There are three possible kinds of states in which a running-wave electron can be: a "spin" state of the TIA assembly, a "spin" state of the TIS assembly and a "full" state. Therefore, there are three corresponding currents of the running-wave electrons, which can be calculated as

$$\vec{j}_{spin,TIS} = -\frac{e \cdot \tau_k}{3} \cdot \int D(E) \cdot |\vec{v}|^2 \cdot \nabla F_{spin,TIS,0} \cdot dE$$

$$\vec{j}_{full} = -\frac{e \cdot \tau_k}{3} \cdot \int D(E) \cdot |\vec{v}|^2 \cdot \nabla F_{full,0} \cdot dE \qquad (34)$$

$$\vec{j}_{spin,TIA} = -\frac{e \cdot \tau_k}{3} \cdot \int D(E) \cdot |\vec{v}|^2 \cdot \nabla F_{spin,TIA,0} \cdot dE$$

All three currents transport the charge, but the spin is only transported by the "spin" states of the TIA assembly. Therefore, the charge and spin currents can be calculated as

$$\vec{j}_{charge} = \vec{j}_{spin,TIS} + \vec{j}_{full} + \vec{j}_{spin,TIA}$$

$$\vec{j}_{spin} = \vec{j}_{spin,TIA} \qquad (35)$$

In the following we calculate the drift spin and charge currents flowing along an electrical field or along a gradient of the chemical potential and the diffusion spin and charge currents flowing along a gradient of spin polarization.

In the case when there is a spatial gradient of the chemical potential $\mu$, the gradient of the distribution function can be calculated as

$$\nabla F_{i,0} = \frac{\partial F_{i,0}}{\partial \mu} \nabla \mu \qquad (36)$$

Substituting Eq. (36) into Eqs. (34) gives

$$\vec{j}_{spin,TIS} = -\nabla \mu \cdot \frac{e \cdot \tau_k}{3} \cdot \int D(E) \cdot |\vec{v}|^2 \cdot \tilde{\sigma}_{spin,TIS} \cdot d\left(\frac{E}{kT}\right) = -\sigma_{spin,TIS} \frac{\nabla \mu}{e}$$

$$\vec{j}_{full} = -\nabla \mu \cdot \frac{e \cdot \tau_k}{3} \cdot \int D(E) \cdot |\vec{v}|^2 \cdot \tilde{\sigma}_{full} \cdot d\left(\frac{E}{kT}\right) = -\sigma_{full} \frac{\nabla \mu}{e} \qquad (37)$$

$$\vec{j}_{spin,TIA} = -\nabla \mu \cdot \frac{e \cdot \tau_k}{3} \cdot \int D(E) \cdot |\vec{v}|^2 \cdot \tilde{\sigma}_{spin,TIA} \cdot d\left(\frac{E}{kT}\right) = -\sigma_{spin,TIA} \frac{\nabla \mu}{e}$$

where the state conductivities $\tilde{\sigma}_{spin,TIS}, \tilde{\sigma}_{full}, \tilde{\sigma}_{spin,TIA}$ are defined as

$$\tilde{\sigma}_{spin,TIS} = -\frac{\partial F_{spin,TIS,0}}{\partial \left(\frac{E}{kT}\right)} \quad \tilde{\sigma}_{full} = -\frac{\partial F_{full,0}}{\partial \left(\frac{E}{kT}\right)} \quad \tilde{\sigma}_{spin,TIA} = -\frac{\partial F_{spin,TIA,0}}{\partial \left(\frac{E}{kT}\right)} \qquad (38)$$

Eq. (38) was simplified using the fact that in an equilibrium distribution functions $F_{i,0}$ depend only on $\frac{E}{kT}$ and the spin polarization *sp*.

In order to calculate the metal conductivities $\sigma_{spin,TIS}, \sigma_{full}, \sigma_{spin,TIA}$ it is necessary to know the density of states D(E) of the metal (Eqs. (37)). The calculation of the unitless state conductivities $\tilde{\sigma}_{spin,TIS}, \tilde{\sigma}_{full}, \tilde{\sigma}_{spin,TIA}$ does not require the knowledge of the density of states. Therefore, even without knowing the density of states of the metal, some conduction properties of the metal can be calculated and analyzed from the state conductivities. It is an advantage of using the state conductivities.

Similarly, in the case when there is a spatial gradient of the spin polarization *sp*, the gradient of the distribution function can be calculated as

$$\nabla F_{i,0} = \frac{\partial F_{i,0}}{\partial sp} \nabla sp \qquad (39)$$

and the currents can be calculates as

$$\vec{j}_{spin,TIS}^{sp} = \nabla sp \cdot \frac{e \cdot \tau_k}{3} \cdot \int D(E) \cdot |\vec{v}|^2 \cdot \tilde{\sigma}_{spin,TIS}^{sp} \cdot d\left(\frac{E}{kT}\right) = \sigma_{spin,TIS}^{sp} \frac{kT}{e} \nabla sp$$

$$\vec{j}_{full}^{sp} = \nabla sp \cdot \frac{e \cdot \tau_k}{3} \cdot \int D(E) \cdot |\vec{v}|^2 \cdot \tilde{\sigma}_{full}^{sp} \cdot d\left(\frac{E}{kT}\right) = \sigma_{full}^{sp} \frac{kT}{e} \nabla sp \qquad (40)$$

$$\vec{j}_{spin,TIA}^{sp} = \nabla sp \cdot \frac{e \cdot \tau_k}{3} \cdot \int D(E) \cdot |\vec{v}|^2 \cdot \tilde{\sigma}_{spin,TIA}^{sp} \cdot d\left(\frac{E}{kT}\right) = \sigma_{spin,TIA}^{sp} \frac{kT}{e} \nabla sp$$

where the state conductivities are defined as

$$\tilde{\sigma}_{spin,TIS}^{sp} = \frac{\partial F_{spin,TIS,0}}{\partial sp} \quad \tilde{\sigma}_{full}^{sp} = \frac{\partial F_{full,0}}{\partial sp} \quad \tilde{\sigma}_{spin,TIA}^{sp} = \frac{\partial F_{spin,TIA,0}}{\partial sp} \qquad (41)$$

Combining Eqs. (37) and (40), the spin and charge currents can be calculated as

$$\vec{j}_{charge} = \vec{j}_{spin,TIS} + \vec{j}_{full} + \vec{j}_{spin,TIA} + \vec{j}_{spin,TIS}^{sp} + \vec{j}_{full}^{sp} + \vec{j}_{spin,TIA}^{sp}$$

$$\vec{j}_{spin} = \vec{j}_{spin,TIA} + \vec{j}_{spin,TIA}^{sp} \qquad (42)$$

or (See Eq. (8))

$$\vec{j}_{charge} = \sigma_{charge} \cdot \frac{1}{e} \nabla \mu + \sigma_{detection} \cdot sp \cdot \frac{kT}{e} \cdot \nabla sp$$

$$\vec{j}_{spin} = \sigma_{injection} \cdot \frac{1}{e} \cdot \nabla \mu + \sigma_{spin} \cdot \frac{kT}{e} \cdot \nabla sp \qquad (43)$$

where the charge, injection, spin and detection conductivities are calculated as

$$\sigma_{charge} = \sigma_{spin,TIA} + \sigma_{spin,TIS} + \sigma_{full}$$

$$\sigma_{injection} = \frac{1}{sp} \cdot \sigma_{spin,TIA}$$

$$\sigma_{detection} = \frac{1}{sp} \cdot \left( \sigma^{sp}_{spin,TIA} + \sigma^{sp}_{spin,TIS} + \sigma^{sp}_{full} \right) \quad (44)$$

$$\sigma_{injection} = \sigma^{sp}_{spin,TIA}$$

The conductivities used in Eqs (44) can be calculated from the corresponding state conductivities as

$$\sigma_i = \frac{e^2 \cdot \tau_k}{3} \cdot \int D(E) \cdot |\vec{v}|^2 \cdot \tilde{\sigma}_i \cdot d\left(\frac{E}{kT}\right) \quad (45)$$

where the state conductivities $\tilde{\sigma}_i$ can be calculated from Eqs. (38) and (40)

Figure 10 shows the state conductivities calculated for the case of an electron transport in the bulk of a conductor. In the bulk of a conductor, the equilibrium distribution functions $F_{spin,TIS,0}, F_{full,0}, F_{spin,TIA,0}$ are described by the spin statistics [4]. It can be assumed that in a conductor with a low density of dislocations, there are no standing-wave electrons (See Chapter 10). All delocalized electrons are the running-wave electrons and all of them contribute to the running-wave electron currents. Therefore, the distribution functions of the spin statistics (See Chapter 4 of Ref. [4]) are used in Eqs. (38), (41) in order to calculate the state conductivities shown in Fig.10. Figure 11 shows the charge, spin-diffusion, injection and detection state conductivities calculated from Eqs. (44) using data obtained for Fig.10.

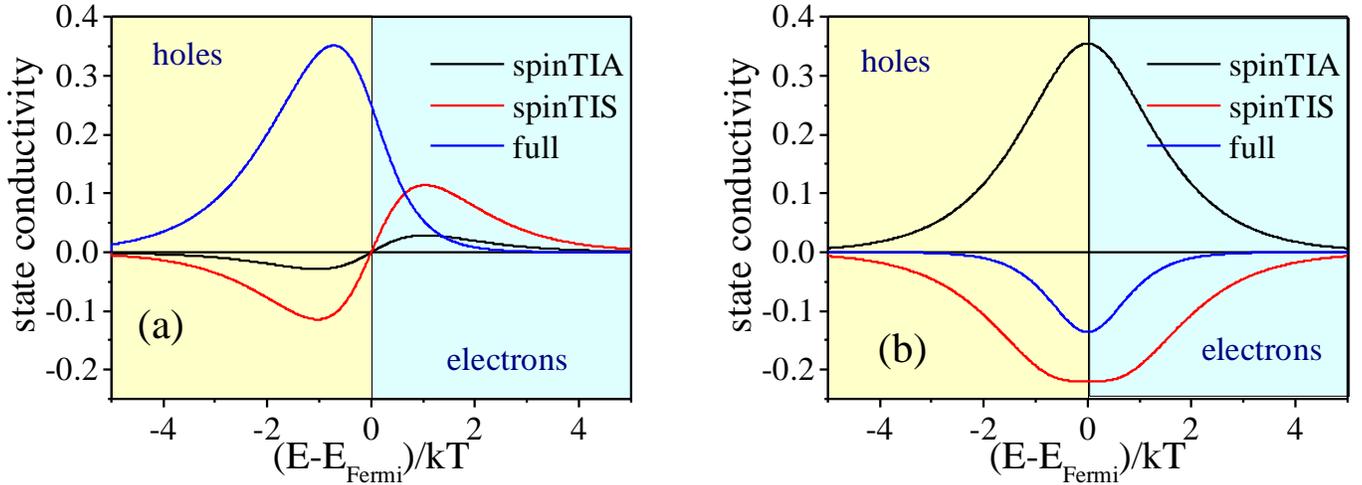

*Fig.10. State conductivities of the "spin" states of the TIA and TIS assemblies and the "full" states of the TIS assembly. Spin polarization of the electron gas is 20 %. The current of electrons of energy larger than the Fermi energy is defined as the electron current (blue area). The current of electrons of energy smaller than the Fermi energy is defined as the hole current (yellow area). (a) $\tilde{\sigma}_{spin,TIS}, \tilde{\sigma}_{full}, \tilde{\sigma}_{spin,TIA}$ (See Eq.38); (b) $\tilde{\sigma}^{sp}_{spin,TIS}, \tilde{\sigma}^{sp}_{full}, \tilde{\sigma}^{sp}_{spin,TIA}$ (See Eq.41).*

Figures 10(a) and 11(a) show the calculated state conductivities, which describe a drift of electrons along a gradient of the chemical potential. The state conductivity $\tilde{\sigma}_{full}$ for a current of "full" states is positive for all energies. This means that the "full" states as negatively-charged particles are drifted from a "–" source toward a "+" drain. The state conductivities $\tilde{\sigma}_{spin,TIS}, \tilde{\sigma}_{spin,TIA}$ for "spin" states are positive for energies above the Fermi energy and they are negative for energies below the Fermi energy. This means that the drift direction of "spin" states depends on their energy. The

"spin" states of energies above the Fermi energy are drifted from a "–" source toward a "+" drain. The "spin" states of energies below the Fermi energy are drifted in the opposite direction from a "+" source toward a "-" drain.

In a non-degenerate n-type semiconductor, the Fermi energy is below the conduction band and all states available for transport are at energies above the Fermi energy. In this case the state conductivity $\tilde{\sigma}_{full}$ of the "full" states is small and the major transport mechanisms is the transport of the "spin" states. A "spin" state is negatively charged and its spin is ½. Therefore, the "spin" states carry simultaneously the spin and the negative charge. It is similar to the movement an electron in vacuum, when the spin and the negative charge are carried in the same direction. For this reason, the conductivity in an n-type semiconductor is called the electron conductivity.

In a non-degenerate p-type semiconductor, the Fermi energy is above the valence band and all states available for transport are at energies below the Fermi energy. In this case both the "full" and "spin" states substantially contribute to the transport. Since $\tilde{\sigma}_{full}$ is positive and $\tilde{\sigma}_{spin}$ is negative, the "full" and "spin" states transport a negative charge in opposite directions. The $\tilde{\sigma}_{full}$ is larger than $\tilde{\sigma}_{spin,TIS} + \tilde{\sigma}_{spin,TIA}$. Therefore, the charge conductivity $\tilde{\sigma}_{charge}$ is positive for all energies (Fig.11, blue curve) and in total the negative charge is transported from "-" to "+". The spin of a "full" state is zero and only "spin" states carry the spin. The spin is drifted in the opposite direction to the drift direction of the negative charge. It is from "+" to "-". It is similar to the movement a positive particle in vacuum, when the spin and the positive charge are carried in the same direction. The conductivity in a p-type semiconductor is called the hole conductivity.

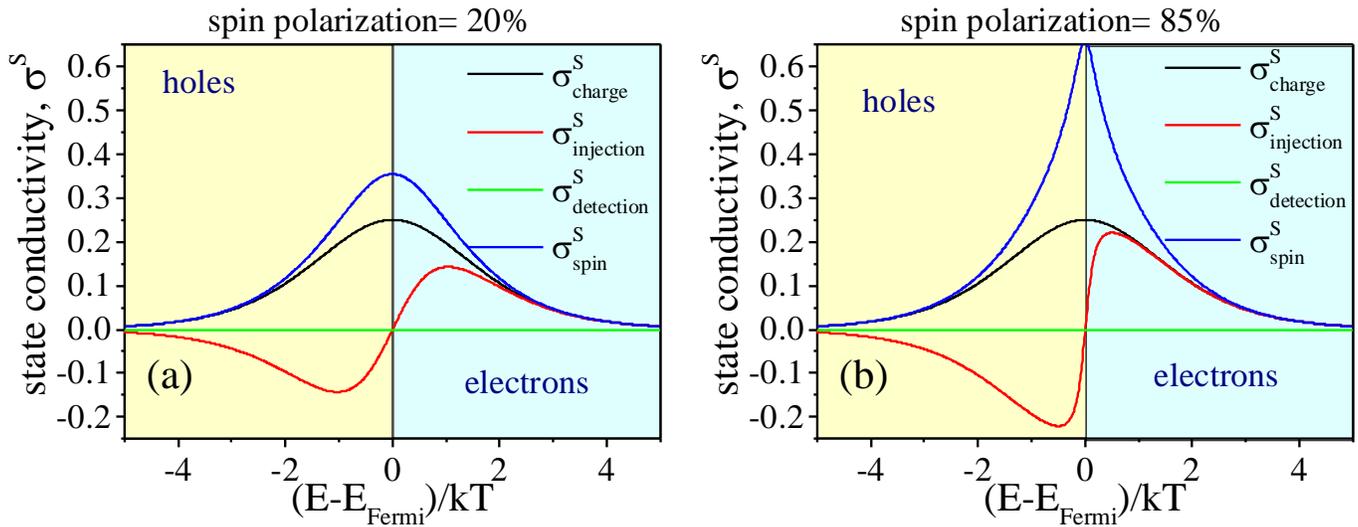

*Fig. 11. The charge, injection, detection and spin-diffusion conductivities for a running wave electron current in the bulk of a conductor with a low density of defects.(a) spin polarization **sp**=0.2 (b) spin polarization **sp**=0.85*

As was explained above (See Fig.7), in a metal the electron and hole conductivities are almost equal. The drift direction of the charge is the same for the electron and hole currents. It is a drift of a negative charge from "-" to "+" or a drift of a positive charge from "+" to "-". In contrast, the drift of spin is in opposite directions for the hole and electron currents and the opposite spin currents nearly compensate each other. Therefore, in a metal the spin drift current is significantly smaller than the drift charge current and the spin polarization of the drift current is always significantly smaller than the spin polarization of the electron gas. It is different from the case of a semiconductor, where the spin polarization of the drift current is equal to the spin polarization of the electron gas.

The above statement can verified from Fig. 11. The injection state conductivity $\sigma^S_{injection}$ is asymmetric with respect to the Fermi energy. The total injection conductivity $\sigma_{injection}$ can be calculated by integrating the injection state conductivity $\sigma^S_{injection}$ over all states (Eq.45). In the case of a non-degenerate semiconductor the states are either above or below the Fermi level. In this case the value of the injection conductivity is the same as the value of the charge conductivity (conventional metal conductivity). The injection conductivity is of opposite sign for n- and p-type semiconductors. In a metal, the density of states is nearly constant at the Fermi energy. The integration (Eq.45) of the product of asymmetric and the nearly-symmetric functions $\sigma^S_{injection}(E) \cdot D(E)$ leads to a small value of $\sigma_{injection}$, which is proportional to the gradient of the density of states at the Fermi energy.

In case when the Fermi energy is far from any band critical point, the density of states is nearly constant and it can be approximate as

$$D(E) \cdot |\vec{v}|^2 = D_F \cdot |\vec{v}_F|^2 + (E - E_F) \cdot \frac{\partial \left( D \cdot |\vec{v}|^2 \right)}{\partial E} \tag{46}$$

where $\vec{v}_F$ and $D_F$ are the electron speed and the density of states at the Fermi energy, respectively.

$$\frac{\partial \left( D \cdot |\vec{v}|^2 \right)}{\partial E} \ll D_0 \tag{47}$$

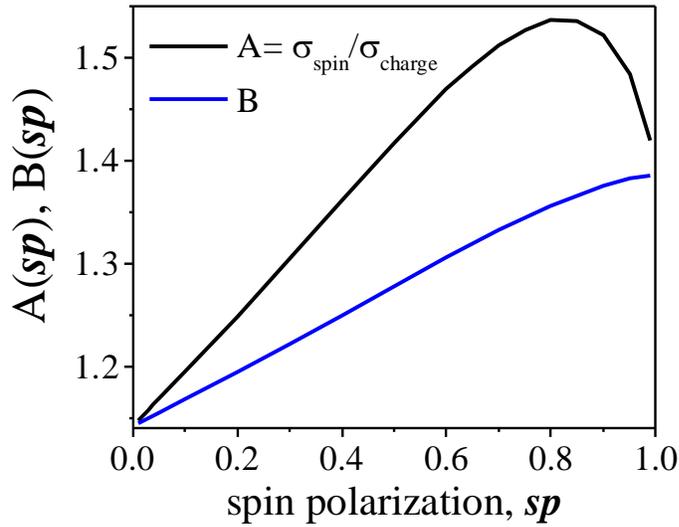

Fig. 12. *The ratio of the spin conductivity to charge conductivity (A(sp)) and the ratio of the injection conductivity to charge conductivity (B(sp)) as functions of the spin polarization of an electron gas (See Eqs. (49))*

Using Eqs. (46) and (47), the integration of Eqs. (45) gives the conductivities in the bulk of a metal with a low density of defects as

$$\begin{aligned}
\sigma_{charge} &= \frac{e^2 \cdot \tau_k}{3} \cdot D_F \cdot |\vec{v}_F|^2 \\
\sigma_{spin} &= A(sp) \cdot \frac{e^2 \cdot \tau_k}{3} \cdot D_F \cdot |\vec{v}_F|^2 \\
\sigma_{injection} &= B(sp) \cdot \frac{e^2 \cdot \tau_k}{3 \cdot kT} \cdot \frac{\partial D}{\partial E} \cdot |\vec{v}_F|^2 \\
\sigma_{detection} &= 0
\end{aligned} \tag{48}$$

where calculated A(*sp*) and B(*sp*) are shown in Fig. 12. From Eqs. (48) the following ratios between the conductivities in the bulk of a metal are obtained

$$\frac{\sigma_{spin}}{\sigma_{charge}} = A(sp)$$

$$\frac{\sigma_{injection}}{\sigma_{charge}} = B(sp) \cdot \frac{kT}{D_F} \cdot \frac{\partial D}{\partial E} \qquad (49)$$

$$\sigma_{detection} = 0$$

In comparison in the case of a non-degenerate semiconductor the conductivity ratios are

$$\frac{\sigma_{spin}}{\sigma_{charge}} = 1 \quad \frac{\sigma_{injection}}{\sigma_{charge}} = \pm 1 \quad \sigma_{detection} = 0 \qquad (50)$$

As was mentioned above, the sign of $\sigma_{injection}$ is opposite for n- and p-type semiconductors. In the case of a metal, the sign of $\sigma_{injection}$ can be either positive or negative and it can be determined from the Hall voltage (See Chapter 12).

Since $\sigma_{charge}$ is the conventional conductivity of metal or semiconductor, which is known or can be measured experimentally, Eqs. (49) and (50) allows to evaluate all the other 3 conductivities. The known values of 4 conductivities should be used in the Spin/Charge Transport equations (Eq. 12).

## *9. Scattering current*

The running-wave-electron conduction occurs because of the electron movements between scatterings. The scattering conduction occurs because of the electron movements during scatterings. In an electron gas the electrons are constantly scattered between quantum states. Because the quantum states have different coordinates in space, each scattering causes an electron spatial movement. In the case when the electron scattering probability is the same in all directions, there is no scattering current. The scattering current can only exists when there is a difference in the electron scattering probability between two opposite directions. For example, the scattering conduction is significant in the vicinity of an interface, because of a substantial difference in the electron scattering probability toward and away from the interface. The hopping conductivity and the Spin Hall effect are examples of the scattering conductivity.

The scattering current is significantly less efficient for the spin and charge transport than the running-wave-electron current. Its contribution to the total current becomes substantial only when the current of running-wave electrons becomes small (for example, the hopping conductivity or the conductivity through a high-resistance contact (a tunnel barrier or a semiconductor-metal contact)) or when the scattering current flows perpendicularly to the current of the running-wave electrons (the Spin Hall effect).

The scattering current occurs because the electrons are scattered from one quantum state to another. According to the type of the quantum states between which electrons are scattered, three types of scattering currents can be distinguished. The electron scattering current is defined as a current where an electron is scattered from a "spin" into an "empty" state (Fig.13 (a)). The hole scattering current is defined as a current where an electron is scattered from a "full" state into a "spin" state (Fig. 13 (b)). The "full"/"empty" scattering current is defined as a current where an electron is scattered from a "full" state into an "empty" state and between two "spin" states (Fig. 13 (c)).

In the following the Boltzmann transport equations are solved for the electron scattering current. The scattering probability from a "spin" state into an "empty" state is proportional to the number of states from which the electron is scattered $F_{spin}$, the number of states to which the electron is scattered $F_{empty}$, the overlap integral of wave functions between which the scattering occurs and the probability that the electron interacts with a defect or phonon or other scattering source. The scattering probability may be different for an electron scattered in the forward direction and for an electron scattered in the backward direction. The reason of this difference can be the difference of the overlap integrals $p_{forward}$ and $p_{backw}$ for scatterings in the forward and a backward directions:

$$p_{forward}(x) = \int \Psi_{spin}(x) \cdot \Psi_{empty}(x+l_{scat}) \cdot dV$$
$$p_{backw}(x) = \int \Psi_{spin}(x) \cdot \Psi_{empty}(x-l_{scat}) \cdot dV \tag{51}$$

where $\Psi_{spin}, \Psi_{empty}$ are wave functions for "spin" and "empty" states, $l_{scat}$ is an average distance through which an electron moves between scatterings.

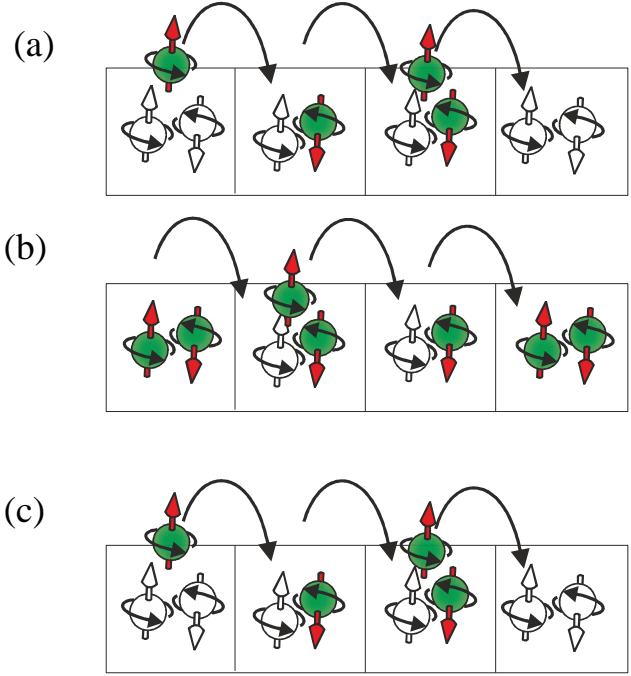

*Fig. 13. Scattering currents. (a) electron current: scattering of an electron from a "spin" state to an "empty" state; (b) hole current: scattering of an electron from a "full" state to a "spin" state; (c) full-empty current: scattering of an electron from a "full" state to an "empty" state (right) and from a "spin" state to a "spin" state (left).*

During time dt an electron moves a distance $dr = |v| \cdot dt$. The mean-free path $\lambda_{mean}$ is an average distance an electron moves between scatterings. Then, the probability for electron to be scattered during time dt is calculated as

$$p_{scattering} = \frac{dr}{\lambda_{mean}} = \frac{|v| \cdot dt}{\lambda_{mean}} \tag{52}$$

Between an "empty" state and a "spin" state, four possible scattering events should be considered. Event 1: scattering of an electron from a "spin" state at position x forward to an "empty" state. Because of this event, the number of spin states at point x decreases

$$dF_{spin}(x) = -p_{forward}(x) \cdot F_{spin}(x) \cdot F_{empty}(x+l_{scat}) \cdot \frac{|v| \cdot dt}{\lambda_{mean,spin}(x)} \tag{53}$$

Event 2: scattering backward from a "spin" state at position x. Because of this event, the number of spin states at point x decreases

$$dF_{spin}(x) = -p_{backw}(x) \cdot F_{spin}(x) \cdot F_{empty}(x-l_{scat}) \cdot \frac{|v| \cdot dt}{\lambda_{mean,spin}(x)} \tag{54}$$

Event 3: scattering backward to an "empty" state at position x from a "spin" state at position x+l$_{scat}$. Because of this event, the number of spin states at point x increases

$$dF_{spin}(x) = p_{backw}(x+l_{scat}) \cdot F_{spin}(x+l_{scat}) \cdot F_{empty}(x) \cdot \frac{|v| \cdot dt}{\lambda_{mean,spin}(x+l_{scat})} \quad (55)$$

Event 4: scattering forward to an "empty" state at position x from a "spin" state at position x-l$_{scat}$. Because of this event, the number of spin states at point x increases

$$dF_{spin}(x) = p_{forward}(x-l_{scat}) \cdot F_{spin}(x-l_{scat}) \cdot F_{empty}(x) \cdot \frac{|v| \cdot dt}{\lambda_{mean,spin}(x-l_{scat})} \quad (56)$$

Summing up all probabilities under the condition that $l_{scat}$ is small, we obtain

$$dF_{spin}(x) = |v| \cdot dt \cdot l_{scat} \frac{\partial}{\partial x}\left[\frac{(p_{backw} - p_{forward}) \cdot F_{spin} \cdot F_{empty}}{\lambda_{mean,spin}}\right] \quad (57)$$

In Eq. (57) it is assumed that during a scattering an electron moves only along the x-direction. In a general case the electron may move also in the y- and z-directions. It can be assumed that during a scattering, an electron moves along its propagation speed. Then, the average distance l$_{scat,x}$ through which an electron moves along the x direction after one scattering is calculated as

$$l_{scat,x} = l_{scat} \cdot \frac{v_x}{|v|} \quad (58)$$

Substituting Eq. (58) into Eq. (57) gives

$$\left(\frac{\partial F_{spin}}{\partial t}\right)_{scattering} = \vec{v} \cdot l_{scat} \cdot \nabla\left((p_{backw} - p_{forward}) \cdot \frac{F_{spin} \cdot F_{empty}}{\lambda_{mean,spin}}\right) \quad (59)$$

Eq. (59) describes the scattering when an electron is scattered from a "spin" state into an "empty" state. Therefore, the "spin" state is the source of the scattering and the "empty" state is the destination. Another type of scattering can occur between a "spin" state and an "empty" state. It is when an "empty" state is the source and a "spin" state is the destination. The second contribution can be calculating similar to Eq. (59) but replacing $\lambda_{mean,spin} \rightarrow \lambda_{mean,empty}$

$$\left(\frac{\partial F_{spin}}{\partial t}\right)_{scattering} = \vec{v} \cdot l_{scat} \cdot \nabla\left((p_{backw} - p_{forward}) \cdot \frac{F_{spin} \cdot F_{empty}}{\lambda_{mean,empty}}\right) \quad (60)$$

Summing up Eqs. (59) and (60) the scattering term of the Boltzmann transport equation can be calculated as

$$\left(\frac{\partial F_{spin}}{\partial t}\right)_{scattering} = \vec{v} \cdot l_{scat} \cdot \nabla\left((p_{backw} - p_{forward}) \cdot F_{spin} \cdot F_{empty}\left(\frac{1}{\lambda_{mean,spin}} + \frac{1}{\lambda_{mean,empty}}\right)\right) \quad (61)$$

The Boltzmann transport equation Eq. (16) only with the scattering term and relaxations term in a static case is simplified to

$$\frac{\partial F_{spin}}{\partial t} = \left(\frac{\partial F_{spin}}{\partial t}\right)_{scattering} + \left(\frac{\partial F_{spin}}{\partial t}\right)_{relaxation} = 0 \quad (62)$$

Using the relaxation-time approximation Eq. (19) and substituting Eq (61) in Eq (62) gives

$$F_{spin,1} = \tau_k \cdot \vec{v} \cdot l_{scat} \cdot \nabla \left[ \left( p_{backw} - p_{forward} \right) \cdot F_{spin,0} \cdot F_{empty,0} \cdot \left( \frac{1}{\lambda_{mean,spin}} + \frac{1}{\lambda_{mean,empty}} \right) \right] \quad (63)$$

Integrating over all possible scattering distances $l_{scat}$, the electron scattering current can be found from Eq.(63) as

$$\vec{j}_{electron} = \frac{-e}{(2\pi\hbar)^3} \iiint \vec{v} \cdot F_{spin,1} \cdot d\vec{p} =$$

$$= -\frac{e \cdot \tau_k}{(2\pi\hbar)^3} \cdot \iiint \cdot d\vec{p} \cdot |\vec{v}|^2 \cdot \int_0^\infty dl_{scat} \cdot l_{scat} \cdot \nabla \left[ \left( p_{backw} - p_{forward} \right) F_{spin,0} \cdot F_{empty,0} \cdot \left( \frac{1}{\lambda_{mean,empty}} + \frac{1}{\lambda_{mean,spin}} \right) \right] \quad (64)$$

## *10. The Mean-Free Path $\lambda_{mean}$*

The mean-free path $\lambda_{mean}$ is the distance through which an electron moves between scatterings. Since an electron is a wave package, the mean-free path can be considered the effective width of this electron wave package or an effective size of an electron or the effective length within which the electron interacts with defects, phonons and other inhomogeneities.

The value of the mean-free path $\lambda_{mean}$ significantly influences the electron transport. In this chapter it is shown the properties of the spin and charge transport changes significantly in the case when in a conductor the mean-free path $\lambda_{mean}$ becomes comparable with an average distance between defects. In next chapter it is shown that at distances of about $\lambda_{mean}$ from an interface even the mechanism the electron transport may change. The mechanism may change from the current of the running-wave electrons in the bulk to the scattering current in the vicinity of the interface.

It is important to note that the mean-free path depends on the electron energy. For example, an electron with an energy sufficiently lower than the Fermi energy has only a few unoccupied states where it can be scattered into. Therefore, the mean-free path of for this electron is long. In contrast, near the Fermi energy there is a sufficient number of unoccupied states, an electron of this energy has a shorter mean-free path.

Three mechanisms, which influence $\lambda_{mean}$, can be distinguished. The first mechanism is the quantum-mechanical limitation for a scattering. An electron is a fermion and it can be scattered only if there is an unoccupied state where the electron can be scattered into. As was mentioned above, an electron of energy substantially lower than the Fermi energy has a very low scattering probability. The second mechanism is the mean-free-path-dependent scattering. This is the scattering on defects and dislocations, which have some spatial distribution with some average distance between them. When $\lambda_{mean}$ is significantly smaller than this distance, the scattering probability is small, but when $\lambda_{mean}$ is comparable or longer than the average distance between defects, the scattering probability becomes larger. The third mechanism is the mean-free-path-independent scattering. This is the scattering on phonons and magnons, which does not depend on $\lambda_{mean}$. Combining these three contributions, the mean-free path $\lambda_{mean}$. can be calculated as

$$\lambda_{mean} = \frac{\lambda_{mean,1}}{p_{quantum} \left( p_{\lambda\_indep} + p_{\lambda\_dep} \right)} = \frac{\lambda_{mean,0}}{p_{quantum} \left( 1 + \frac{p_{\lambda\_dep}}{p_{\lambda\_indep}} \right)} \quad (65)$$

where $p_{quantum}$ is the probability that there is an available unoccupied state into which an electron can be scattered, $p_{\lambda\_indep}$ is the scattering probability, which is independent of the mean free path and $p_{\lambda\_dep}$ is the scattering probability, which is dependents on the mean-free path.; $\lambda_{mean,0}$ is the mean-free path in the case when there are no mean-free-path dependent scatterings and there is a sufficient number of states which an electron can be scattered into: $p_{quantum} = 1 \quad p_{\lambda\_dep} = 0$

The distribution of the defects in a metal depends on the metal fabrication method. The normal or Gaussian distribution of defects is the most probable. In this case the probability, that the distance from one defect to the closest neighbor defect equals d, can be calculated as

$$p_{2defects}(d) = A \cdot \exp\left(-\frac{(d-d_0)^2}{\Delta d^2}\right) \qquad (66)$$

where $d_0$ is the average distance between defects, $\Delta d$ is the width of the distribution and

$$A = \frac{2}{\Delta d \cdot \sqrt{\pi} \cdot \left(1 + Erf\left(\frac{d_0}{\Delta d}\right)\right)}$$

Next, it can be assumed that an electron can be scattered only if its mean-free path $\lambda_{mean}$ is larger than the distance between two defects d. Then, the probability $p_{\lambda\_dep}$ for a mean-free-path dependent scattering is calculated as

$$p_{\lambda\_dep}(\lambda_{mean}) = \int_0^{\lambda_{mean}} p_{2defects}(x) \cdot dx \qquad (67)$$

Substituting Eq. (62) into Eq. (61) and integrating gives the probability $p_{\lambda\_dep}$ for a mean-free-path dependent scattering as

$$p_{\lambda\_dep}(\lambda_{mean}) = \frac{Erf\left(\frac{\lambda-d_0}{\Delta d}\right) + Erf\left(\frac{d_0}{\Delta d}\right)}{1 + Erf\left(\frac{d_0}{\Delta d}\right)} \qquad (68)$$

$p_{quantum}$ can be calculated as follows. The scattering probability is proportional to the number of unoccupied states.

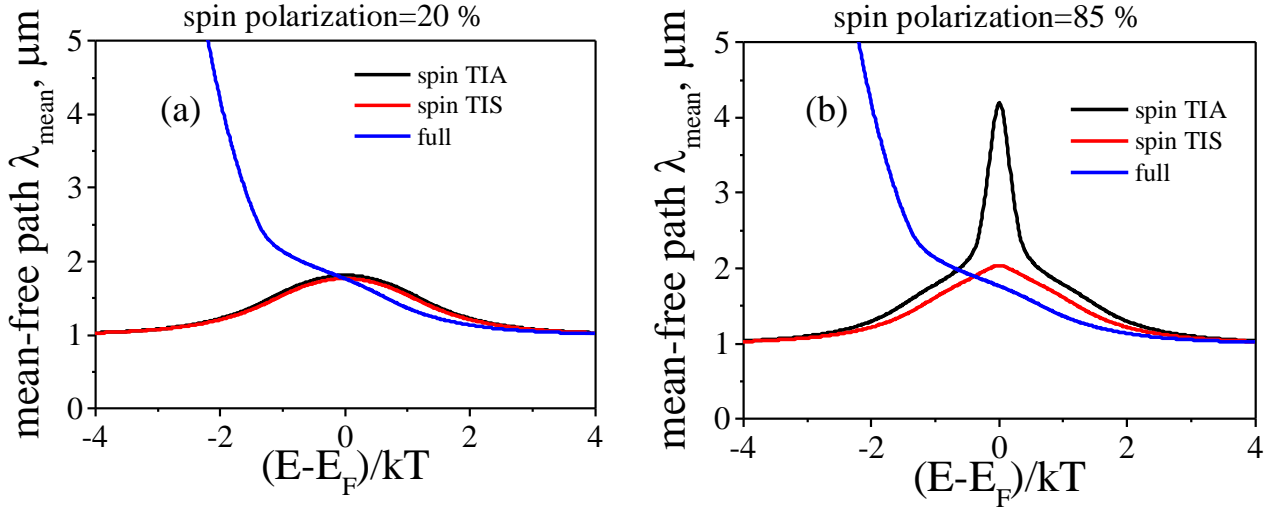

*Fig. 14. Mean-free path $\lambda_{mean}$ of the "spin" states of the TIA assembly and the "full" and "spin" states of the TIS assembly (a) spin polarization **sp**=0.2 (8) spin polarization **sp**=0.85*

Therefore, $p_{quantum}$ is different for different kinds of states. An electron from a "full" state can be scattered either into an "empty" state or into a "spin" state. The probability that one of these states will be available for a scattering is

$$p_{quantum,full} = F_{empty} + F_{spin,TIS} + F_{spin,TIA} \qquad (69)$$

An electron from a "spin" state can be scattered either into an "empty" state or a "spin" state. From a "spin" state into a "spin" state an electron can be scattered only if the "spin" states have opposite spin directions. Since all electrons of the TIA assembly have the same direction of spin, there are no scatterings between the "spin" states of the TIA assembly. The spin direction of a "spin" state of the TIS assembly may be in any direction with the equal probability. Therefore, the probability of a scattering between a "spin" state of the TIS assembly and a "spin" state of the TIS or TIA assembly equals 1/2 (See Eq. (4) of Ref. [4]). This leads to

$$p_{quantum,spin,TIA} = F_{empty} + F_{full} + \frac{1}{2} F_{spin,TIS}$$

$$p_{quantum,spin,TIS} = F_{empty} + F_{full} + \frac{1}{2}\left(F_{spin,TIS} + F_{spin,TIA}\right)$$

(70)

Substituting Eqs. (68)-(70) into Eq. (66) and solving it numerically, the mean-free path was calculated. Figure 14 shows the mean-free path for "full" states and "spin" states of the TIS and TIA assemblies as a function of energy. The ratio of probabilities of $\lambda_{mean}$-dependent to $\lambda_{mean}$-independent scatterings $\frac{p_{\lambda\_dep}}{p_{\lambda\_indep}}=1$, $\lambda_{mean,0}=1$ μm, the defect reflectivity $R_d=0.8$, the average distance between defects d is 1 μm and Δd is 1 μm. The mean-free path depends significantly on the electron energy. The mean-free path $\lambda_{mean}$ of the "full" states at energies lower than 2kT becomes several times longer than $\lambda_{mean}$ at energies above the Fermi energy. The mean-free path becomes substantially longer, because at these energies there are only a few unoccupied states (a few of "empty" and "spin" states) where an electron can be scattered into. The mean-free path of the "spin" states of the TIA assembly becomes longer at the Fermi energy, when the spin polarization of the electron gas is high. An electron can be scattered into/out of a "spin" state of the TIA assembly only out/into a "spin" state of the TIS assembly or out of a "full" state or into an "empty" state. At the Fermi energy there are few of these states when the spin polarization of the electron gas is high (Fig.5 (b) of Ref. [4]) and all electrons are in the "spin" states of the TIA assembly. Therefore, an electron of a "spin" state of the TIA assembly has only few states where it can be scattered to.

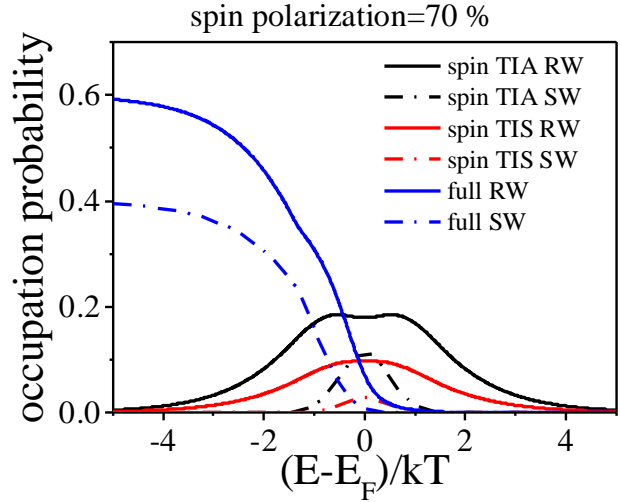

*Fig. 15. Occupation probability of running-wave electrons (RW) and standing-wave electrons (SW) for "full" states and "spin" states of the TIA and TIS assemblies in the bulk of a conductor with defects.*

Figure 15 shows the occupation probability for the running-wave electrons and the standing-wave electrons. The largest number of standing-wave electrons is at energies below the Fermi energy occupying the "full" states. There is a substantial number of standing-wave electrons at the Fermi energy occupying the "spin" states of the TIA assembly.

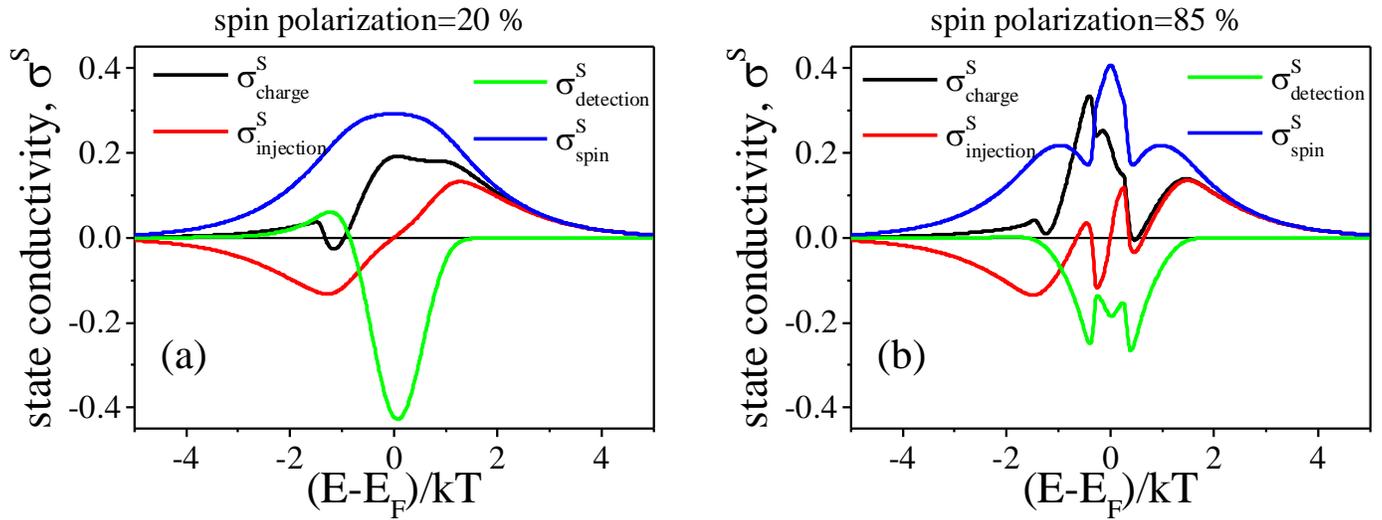

*Fig. 16. The charge, injection, detection and spin-diffusion conductivities for a running wave electron current in the bulk of a conductor with defects. (a) spin polarization **sp**=0.2 (b) spin polarization **sp**=0.85.*

Figure 16 shows the calculated state conductivities for a conductor with defects. The material parameters are the same as were used for Figs. 14 and 15. The state conductivities of Fig. 16 can be compared with the state conductivities in a conductor without defects (See Fig.11). There are two distinguished differences between conductivities in conductors with and without defects. The first difference is the value of the detection conductivity $\sigma_{detection}$. It is zero in a conductor without defects for any electron energy and it is non-zero in a conductor with defects at energies near the Fermi energy. The second difference is the dependence of the charge conductivity $\sigma_{charge}$ on the spin polarization *sp* of the electron gas. The charge conductivity is independent of the spin polarization in a conductor without defects. This means that there is no magneto-resistance effect in a conductor without defects. In contrast, in a conductor with defects, the charge conductivity is dependent on the spin polarization of the electron gas and there is a magneto-resistance effect. The conduction of electrons of only energies near the Fermi energy depends on the spin polarization. Therefore, this magneto-resistance effect may exists only in a metal, but not in a non-degenerate semiconductor.

These two features are explained as follows. As was shown in Chapter 4, in the bulk of a conductor without defects a spin diffusion current flows without any charge diffusion current. Such spin current is possible only because in the bulk of a conductor without defects there is a balance. The numbers of spin-polarized and spin-unpolarized electrons, which diffuse in the opposite directions along the gradient of the spin accumulation, are exactly equal. This results in a spin current without any charge accumulation and $\sigma_{detection}$=0. In a conductor with defects the number of the running-wave electrons in "full" states of the TIS assembly (spin-unpolarized electrons) becomes substantially smaller at lower energies. The number of the running-wave electrons in the "spin" states of the TIA assembly (spin-polarized electrons) becomes smaller at the Fermi energy. This unbalances the number of spin-polarized and spin–unpolarized electrons, which diffuse in opposite directions. As a result, the charge is accumulated along the spin diffusion and $\sigma_{detection} \neq 0$.

The reason for the appearance of the magneto-resistance in the bulk of a conductor with defects is similar. In a conductor without defects, all electrons are running-wave electrons. When the spin polarization of the electron gas changes, the numbers of spin-polarized and spin-unpolarized electrons change. However, the total number of the running-wave electrons remains constant and the charge conductivity remains unchangeable. In a conductor with defects the ratio of the standing-wave to running-wave electrons is different for the TIS and TIA assemblies. This means that when the spin polarization of the electron gas changes, the total number of the running-wave electrons changes. As a consequence, the conductivity of the running-wave electrons become dependent on the spin polarization of the electron gas.

It should be noted that the values of the detection conductivity, the injection conductivity and the magneto-resistance depend on a defect distribution. Materials having the larger detection conductivity or/and the larger injection conductivity or/and the larger magneto-resistance are required for a variety of the practical applications. It might be possible that the spin properties of a metal can be engineered by fabricating artificial defects inside the metal with an optimized distribution. A similar technique was proved to be effective in the case of a photonic crystal [20], where the optical properties of the crystal can be engineered by inserting artificial defects into the crystal.

## 11. Transport in the vicinity of an interface

The spin and charge transport in the vicinity of a contact between two metals are substantially different from the transport in the bulk of metals. In the vicinity of the contact, the number of the running-wave electrons decreases and the number of the standing-wave electrons increases compared to the bulk. In the bulk the major transport mechanism is the current of the running-wave electrons and the scattering current is often significantly smaller. In contrast, in the vicinity of an interface the current of the running-wave electrons decreases or even becomes negligible, but the scattering current increases and often became the dominant transport mechanism.

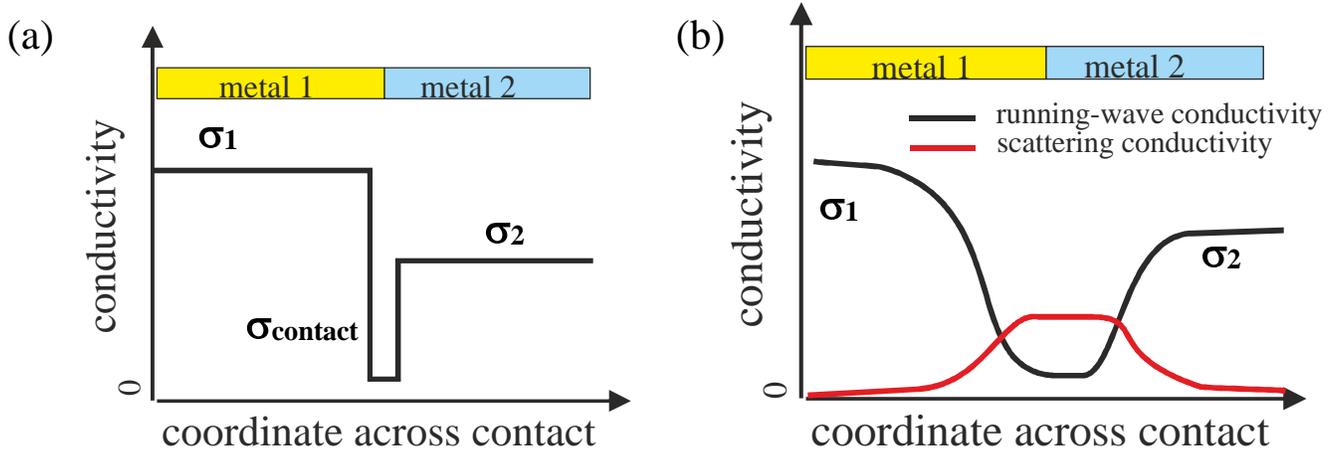

*Fig. 17. Conductivity of a metal in the vicinity of an interface. (a) The classical model. Up to the contact the metal resistance is the same as it is in the bulk. At the interface there is a delta-function-like sharp increase of the resistance, which is called the contact resistance. (b) Presented model. The running-wave electron conductivity decreases from the bulk value of metal 1, has a minimum at the interface and increases to the bulk value of metal 2. The scattering conductivity is small in the bulk, it increases in the vicinity of the contact and it becomes largest at the contact interface.*

Figure 17 compares the conductivity in the vicinity of an interface as it is assumed in the classical model of the electron gas and as it is calculated in the presented model. In the classical model it is assumed that the conductivity at contact is step-like (Fig.17 (a)). Till the contact the conductivity is constant and it equals to the conductivity in the bulk of the metal. On the other side of the contact, the conductivity equals to the bulk conductivity of the second metal. Also, it is assumed that at an interface between metals there is a delta-function-like sharp increase of the resistance, which is called the contact resistance. According to the presented model the conductivity through the contact does not have any step-like or delta-function-like features. The conductivity monolithically changes over a distances ~1-4 mean-free paths from the bulk conductivity of one metal to the bulk conductivity of the other metal. The running-wave conductivity decreases from the bulk conductivity of the first metal, reaches a minimum at the interface between the metals and again increases to the bulk conductivity of the second metal (Fig.17 (b)). The scattering conductivity changes from being negligibly small in the bulk of the metals to becoming a substantial at the interface. The scattering conductivity may even overcome the running-wave conductivity.

A contact or an interface between two metals can be considered is an obstacle for the movement of an electron. The electrons can be compared with photons of light. When light passes through the boundary between two materials, only a part of light passes through the boundary and a part of light is reflected back. Similarly for an electron, there is a finite probability that the electron is reflected from an interface and there is a finite probability the electron passes through the interface as a running-wave electron. Obviously, the electrons, which are reflected from the interface, do not transport the spin and charge through the contact. They can be considered as bounded by the interface and they become the standing-wave electrons. In the case of a contact between similar metals, the electron reflection at the contact is a small and in vicinity of contact there is a substantial number of the running-wave electrons. In contrast, the reflection is nearly full at a contact between different materials, like a metal-semiconductor or a tunnel barrier, and almost all the electrons in the vicinity of this contact are the standing-wave electrons.

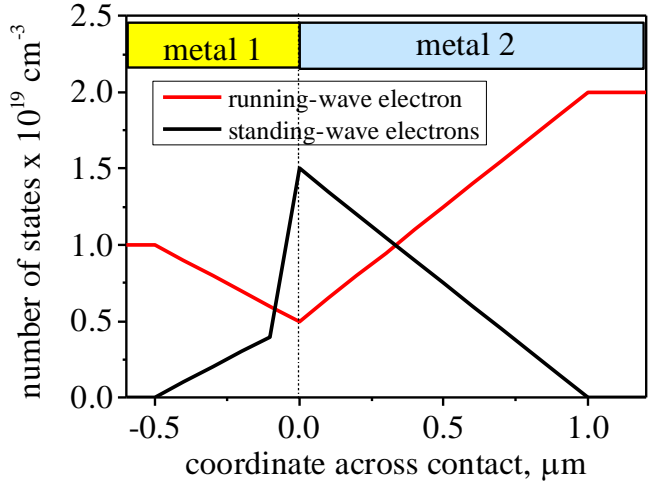

*Fig.18. Number of the running-wave and standing-wave electrons in the vicinity of a contact between two metals. It is assumed that in each metal the mean-free path is the same for all electrons.*

In the following the conductivity of the running-wave electrons is calculated in the vicinity of the interface. The approximation of the square-shape wave-function is used where the special distribution wave function of an electron moving along the x-direction at point $x=x_0$ is approximated as

$$\Psi(x) = \begin{cases} 0 & x < x_0 - \dfrac{\lambda_{mean}}{2} \\ \dfrac{1}{\sqrt{\lambda_{mean}}} & x_0 - \dfrac{\lambda_{mean}}{2} < x < x_0 + \dfrac{\lambda_{mean}}{2} \\ 0 & x > x_0 + \dfrac{\lambda_{mean}}{2} \end{cases} \tag{71}$$

It is assumed that when the wave function of an electron (Eq. (71)) touches the interface, a running-wave electron becomes a standing-wave electron with the probability $R$, where $R$ is defined as the reflectivity of the interface. In the following we approximate that reflectivity of the interface does not depend on an electron incident angle. Even though it is a rather rough approximation, it still gives a clearer description of the features of the conduction near a contact or an interface. The probability for an electron (Eq. (71)) to touch the interface depends on the distance of the electron to the interface and the electron propagation direction with respect to the normal to the interface. In the case when the contact is in the yz-plane at x=0, the electron at position x=x1 touches the contact if

$$x_1 < \frac{\lambda_{mean}}{2} \cdot \cos(\theta) \tag{72}$$

where θ is the angle between the contact normal and the electron propagation direction.
In the case when the distribution of electrons is isotropic, the probability that an electron propagates in the direction within angles between θ and θ+dθ is

$$p(\theta) = 0.5 \cdot \sin(\theta) \cdot d\theta \tag{73}$$

Eqs. (72) and (73) give the probability that the electron wave function (71) touches the contact interface as

$$p_{touching} = 2 \cdot \int_0^{a\cos\left(\frac{2 \cdot x_1}{\lambda_{mean}}\right)} 0.5 \cdot \sin(\theta) \cdot d\theta = 1 - \frac{2 \cdot x_1}{\lambda_{mean}} \tag{74}$$

From Eq. (74) the probability for an electron to be a standing-wave electron at distance $x_1$ from the contact interface is calculated as

$$p_{standing}(x_1) = \begin{cases} R \cdot \left(1 - \dfrac{2 \cdot x_1}{\lambda_{mean}}\right) & x_1 < \lambda_{mean}/2 \\ 0 & x_1 > \lambda_{mean}/2 \end{cases} \tag{75}$$

It should be noted that in the case of a more realistic Gaussian-shape wave function (Gaussian-shape envelop function), the probability of electron to be a standing-wave electron depends on the amount of overlap of the electron wave function with the regions of the first and second metals
.

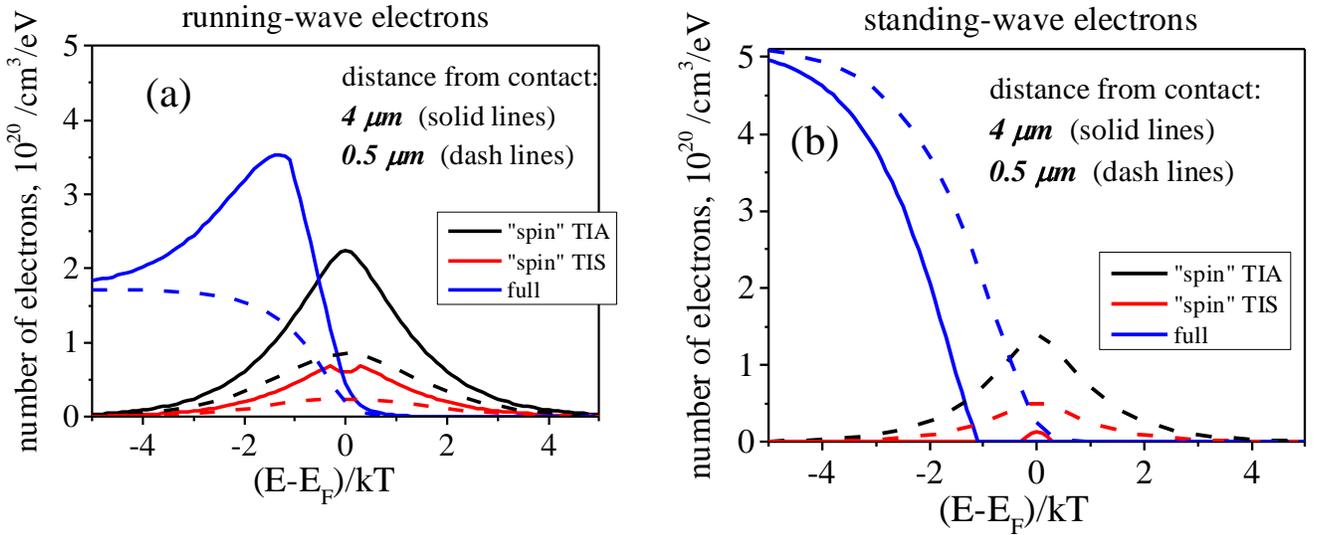

*Fig. 19. (a) Number of running-wave electrons and (b) standing-wave electrons in metal 2 at distances 4 μm (solid lines) and 0.5 μm (dash lines) from the contact with metals 1.*

When an electron is passing through a contact, it can be reflected or scattered to become a standing-wave electron. Otherwise, it is passing the contact as a running-wave electron. In the vicinity of a contact, the electron wave function spreads into both metals. As a running-wave electron approaches the contact interface, the more of its wave function penetrates into the second metal making the probability of the electron reflection higher. This is the reason of the continuous decrease of the number of the running-wave electron when the distance to interface decreases. As a running-wave electron moves through the contact interface, the overlap of its wave function with the regions of the first and second metals changes continuously without a step. This literally means that at both sides of the contact the number of the running-wave electrons should be the same or in other words the distribution of the running-wave electrons throughout the contact should be without steps. This condition gives the ratio between contact reflectivity $R_1$ of metal 1 and contact reflectivity $R_2$ of metal 2 as

$$n_1 \cdot (1 - R_1) = n_2 \cdot (1 - R_2) \tag{76}$$

where n1 and n2 are the number of the running-wave electrons at the contact interface in metals 1 and 2, respectively. From Eqs. (75) and (76) the numbers of the running-wave electrons and the standing-wave electrons are calculated as

$$n_{running}(x) = \begin{cases} n_1 & x < -\lambda_{mean,1}/2 \\ n_1 \cdot \left[1 - R_1 \cdot \left(1 + \dfrac{2 \cdot x}{\lambda_{mean,1}}\right)\right] & -\lambda_{mean,1}/2 < x < 0 \\ n_2 \cdot \left[1 - R_2 \cdot \left(1 - \dfrac{2 \cdot x}{\lambda_{mean,2}}\right)\right] & 0 < x < \lambda_{mean,2}/2 \\ n_2 & x > \lambda_{mean,2}/2 \end{cases} \qquad (77)$$

$$n_{standing}(x) = \begin{cases} 0 & x < -\lambda_{mean,1}/2 \\ n_1 \cdot R_1 \cdot \left(1 + \dfrac{2 \cdot x}{\lambda_{mean,1}}\right) & -\lambda_{mean,1}/2 < x < 0 \\ n_2 \cdot R_2 \cdot \left(1 - \dfrac{2 \cdot x}{\lambda_{mean,2}}\right) & 0 < x < \lambda_{mean,2}/2 \\ 0 & x > \lambda_{mean,2}/2 \end{cases} \qquad (78)$$

Figure 18 shows the distribution of running-wave and standing-wave electrons in the vicinity of a contact between two metals with different numbers of the conduction electrons. The number of running-wave electrons is 1E19 1/cm$^3$ and 2E19 1/cm$^3$ in the bulk of the metals 1 and 2, respectively. The mean-free path is 1 μm and 2 μm in metals 1 and 2, respectively. To make easier to understand the influence of the interface, it is assumed that in each metal the mean-free path is the same for all electrons. The reflectivity of the contact is 50% for the metal 1 and 75% for the metal 2. At a distances longer than $\lambda_{mean}$ /2 from the interface all electrons are running-wave electrons. For shorter distances the number of the running-wave electrons decreases becoming smallest at the interface. In contrast, the number of the standing-wave electrons increases becoming largest at the interface.

For calculation of Fig. 18 it is assumed that in each metal the mean-free path is the same for all electrons. However, in the previous chapter it was shown that the mean-free path depends significantly on the electron energy and the type of the state the electron occupies. Figure 19 shows the calculated number of "full" states and "spin" states of the TIA and the TIS assemblies in metal 2 at distances 4 μm and 0.5 μm from the contact with metals 1. The energy dependence of that $\lambda_{mean}$ was included into the calculations. The contact parameters are the same as were used for Fig. 18. The spin polarization of both metals is 75%. It is assumed that in both metals there are no defects and all scatterings are $\lambda_{mean}$ - independent.

The "full" states are the most affected in the proximity of the contact interface. Even at the long distance of 4 μm from the contact there is a significant number of the standing-wave electrons occupying the "full" states at low energy. For a shorter distance of 0.5 μm there is a significant number the standing-wave electrons of all types and all energies.

As can be seen from Eq. (72) the number of the running-wave electrons depends on the electron moving direction. The closer the moving direction is to the interface normal, the higher the probability for the electron to be the standing-wave type. This makes the conductivity of the running-wave electrons to be anisotropic in the vicinity of a contact. This means that near an interface the conductivity to become different for the currents flowing along and across the interface.

The current that flows perpendicularly to the interface can be calculated from Eq. (33), where θ is both the angle between the electron movement direction and the normal to the interface and the angle between the electron movement direction and the current direction. From Eq. (72) the probability that an electron is of the running-wave type is

$$p_{running}(\theta) = \begin{cases} 1 & a\cos\left(\dfrac{2 \cdot x_1}{\lambda_{mean}}\right) < \theta < \pi/2 \\ 1-R & 0 < \theta < a\cos\left(\dfrac{2 \cdot x_1}{\lambda_{mean}}\right) \end{cases} \qquad (79)$$

Substituting Eq. (79) into Eq. (33) and integrating over all possible θ gives the current flowing perpendicularly to the interface as

$$\vec{j}_\perp(x_1) = -\tau_k e \cdot \int 0.5 \cdot D(E) \cdot dE \cdot |\vec{v}|^2 \cdot \nabla F_{i,0} \cdot \begin{cases} \int_{a\cos\left(\frac{2 \cdot x_1}{\lambda_{mean}}\right)}^{\pi/2} 2\cdot\sin(\theta)\cdot\cos^2(\theta)d\theta + (1-R)\cdot \int_0^{a\cos\left(\frac{2 \cdot x_1}{\lambda_{mean}}\right)} 2\cdot\sin(\theta)\cdot\cos^2(\theta)d\theta & x_1 < \dfrac{\lambda_{mean}}{2} \\ \int_0^{\pi/2} \sin(\theta)\cdot\cos^2(\theta)d\theta & x_1 \geq \dfrac{\lambda_{mean}}{2} \end{cases}$$

$$= -\dfrac{\tau_k e}{3} \cdot \int D(E)\cdot dE \cdot |\vec{v}|^2 \cdot \nabla F_{i,0} \begin{cases} R\cdot\left(\dfrac{2\cdot x_1}{\lambda_{mean}}\right)^3 + 1 - R & x_1 < \dfrac{\lambda_{mean}}{2} \\ 1 & x_1 \geq \dfrac{\lambda_{mean}}{2} \end{cases} \qquad (80)$$

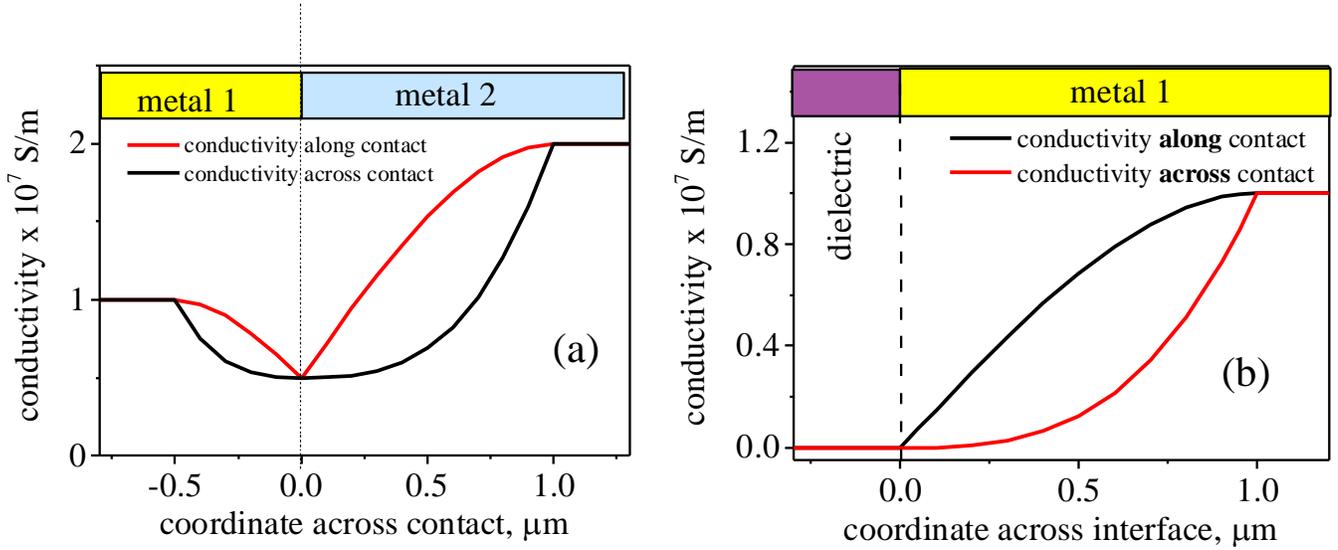

*Fig. 20. Conductivity in the direction perpendicular to the contact (black line) and parallel to the contact (red line) in the vicinity of a contact between (a) two metals and (b) a metal and a dielectric. It is assumed that in each metal the mean-free path is the same for all electrons.*

In order to calculate the current flowing along the interface, we use the cylindrical coordinates, where θ is defined as the angle between the electron movement direction and the direction of current flow. ϕ is the angle between the normal to the interface and the projection of the electron movement direction into the plane perpendicular to the current flow. Than, the condition that the electron of the running wave-type is

$$x_1 > \dfrac{\lambda_{mean}}{2}\cdot\sin(\theta)\cdot\cos(\phi) \qquad (82)$$

The range of angles, where condition (82) can be satisfied is

$$a\sin\left(\dfrac{2x_1}{\lambda_{mean}}\right) < \theta < \dfrac{\pi}{2} \qquad (83)$$

In the region in the vicinity of the interface, where $x_1 > \frac{\lambda_{mean}}{2} \cdot \sin(\theta) \cdot \cos(\phi)$, the current along the interface can be calculated from Eq. (33), (82), (83) as

$$\vec{j}_\parallel(x_1) = -\tau_k e \cdot \int D(E) \cdot dE \cdot |\vec{v}|^2 \cdot \nabla F_{i,0} \cdot \left\{ \int_0^\pi 0.5 \cdot \sin(\theta) \cdot \cos^2(\theta) d\theta - R \cdot 2 \cdot \int_{a\sin\left(\frac{2 \cdot x_1}{\lambda_{mean}}\right)}^{\pi/2} 0.5 \cdot \sin(\theta) \cdot \cos^2(\theta) d\theta \frac{1}{\pi} \int_{-a\cos\left(\frac{2 \cdot x_1}{\lambda_{mean} \cdot \sin(\theta)}\right)}^{a\cos\left(\frac{2 \cdot x_1}{\lambda_{mean} \cdot \sin(\theta)}\right)} d\phi \right\} = $$

$$= -\frac{\tau_k e}{3} \cdot \int D(E) \cdot dE \cdot |\vec{v}|^2 \cdot \nabla F_{i,0} \cdot \left[ 1 - R \cdot \int_{a\sin\left(\frac{2 \cdot x_1}{\lambda_{mean}}\right)}^{\pi/2} \frac{6}{\pi} \cdot \sin(\theta) \cdot \cos^2(\theta) a\cos\left(\frac{2 \cdot x_1}{\lambda_{mean} \cdot \sin(\theta)}\right) \cdot d\theta \right]$$

(84)

The integral of Eq. (84) can be evaluated numerically. Figure 20 shows the conductivity of the current of the running-wave electrons in the vicinity of a contact between two metals and a contact between a metal and a dielectric. The bulk conductivities of metal 1 and metal 2 are 1E7 S/m and 2E7 S/m, respectively. Other contact parameters are the same as were used for the calculation of Figs. 18 and 19. Through the contact the conductivity decreases from the value of the bulk conductivity of metal 1, reaches a minimum at the contact interface and then increases to the value of the bulk conductivity of metal 2. In the case when the reflectivity R of the contact is 1 (for example, a metal-dielectric contact (Fig.20 (b))), at the contact interface the conductivity of the current of the running-wave electrons becomes zero and the scattering current becomes only the remaining transport mechanism. It is the case, for example, of transport through a tunnel junction. The reduction of the conductivity occurs in each metal within $\lambda_{mean}/2$. The conductivity for a current across the interface decreases sharper than for a current along the interface.

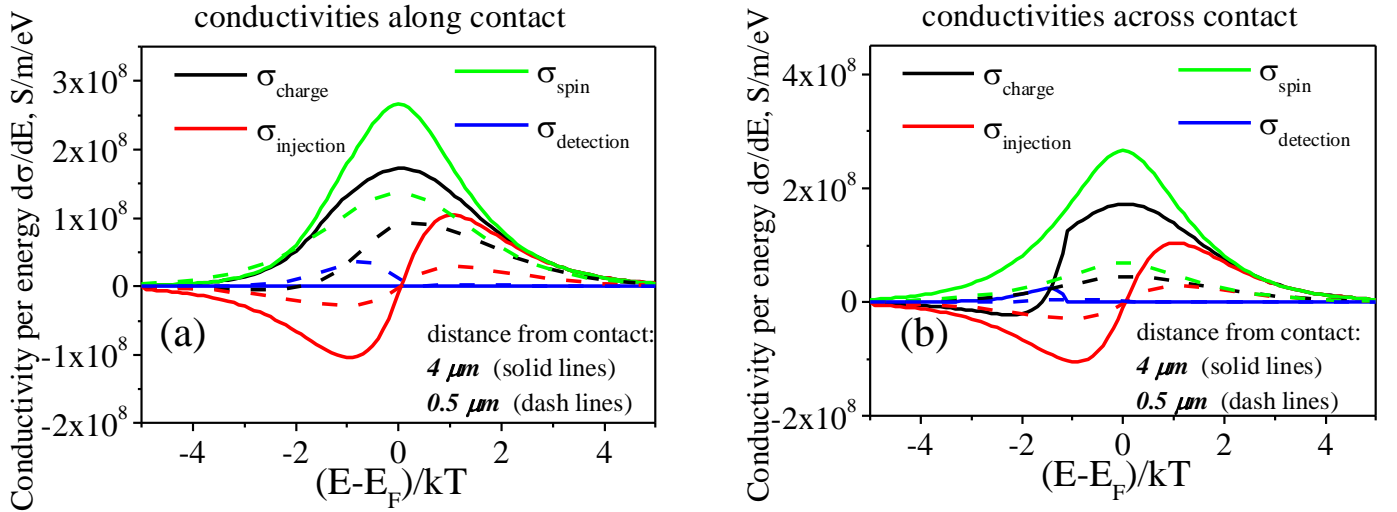

*Fig. 21 (a) charge, injection, spin-diffusion and detection conductivities for a current of running-wave electrons flowing (a) along a contact and (b) across a contact. Solid and dash lines correspond to conductivities in metal 2 at distances of 4 μm and 0.5 μm from contact, respectively.*

For the calculation of Fig. 20 it is assumed that in each metal the mean-free path is the same for all electrons. As was explained above, it is necessary to include in calculation the dependence of $\lambda_{mean}$ on the electron energy and on the type of state, which the electron occupies. Figure 21 shows the charge, injection, spin-diffusion and detection conductivities at distance 4 μm and 0.5 μm from the contact. The contact parameters are the same as were used for the calculation of Figs. 18 -20. Since the density of states may be different for the metals at the sides of the contact, the state conductivities can not describe the change of the conductivity through contact. Instead the derivation of the conductivities $\frac{\partial \sigma_i}{\partial E}$ has been used in Fig.21. The conductivities can be calculated as

$$\sigma_i = \int \frac{\partial \sigma_i}{\partial E} dE \tag{85}$$

Even at the distance of 4 µm from the contact interface, which is longer than the mean-free path $\lambda_{mean,0}$ =2 µm, the conductivities are different from the bulk conductivities, because of the longer mean-free path of the "full" states. For shorter distances the charge and spin conductivities significantly decrease and become different for currents flowing along and across the interface. The changes in the vicinity of the interface are most apparent for the detection conductivity. The detection conductivity is zero in the bulk of a conductor, but it becomes substantial in the vicinity of the interface. As was explained above, the detection conductivity in zero in the bulk, because of the balance of equal flows of spin-polarized and spin-polarized electrons in opposite directions along a gradient of spin accumulation. In the vicinity of the interface this balance is broken.

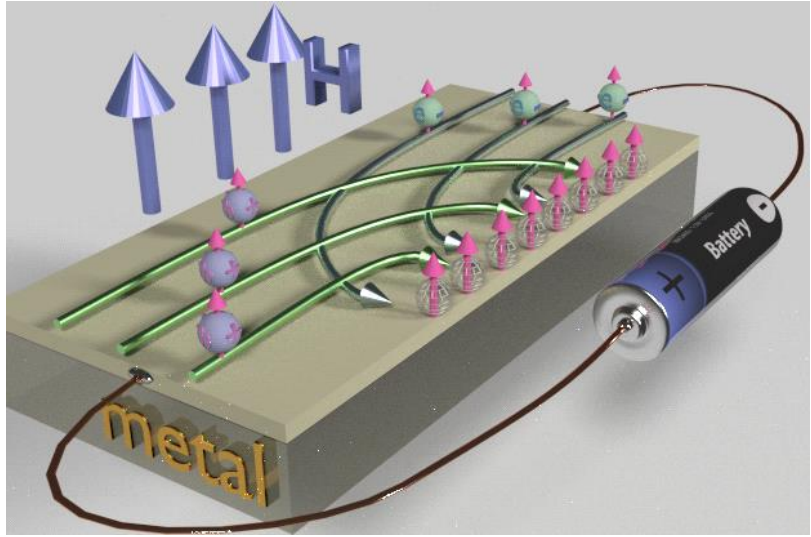

*Fig. 22. The Hall effect in a metal. The electron current (green balls) flows from "-" to "+". In a magnetic field the electrons turn left with respect to the direction of their movement. The hole current (blue balls) flows from "+" to "-". In a magnetic field the holes turn right with respect to the direction of their movement. Both the holes and electrons are accumulated at the same side of the sample. At this side the spins are accumulated, but not the charge.*

## *12. Hall effect.*

An electron experiences several effects in a magnetic field. The first effect is the precession of the electron spin in the magnetic field. There is a damping of the spin precession, which aligns the spin direction along the direction of the magnetic field. Because of the latter effect, a magnetic field induces a spin accumulation in an electron gas [4]. When an electron moves in a magnetic field, it experiences the Lorentz force. The Lorentz force has a relativistic origin. The Theory of Relativity states that a particle moving in a magnetic field experiences an effective electrical field, which is directed perpendicularly to the magnetic field and perpendicularly to the particle movement direction. It is important to emphasize that the direction and magnitude of the effective electrical field does not depend either on the particle charge or on the particle spin.

In the case when a magnetic field is applied perpendicularly to the flow direction of a drift current, the charge carriers experience the Lorentz force in the direction perpendicular to both the magnetic field and the current. Because of this force, the carriers are accumulated at the edges of the sample and a voltage transverse to the drift current is built up. This voltage is called the Hall voltage. In n- and p-type semiconductors the Hall voltage is of opposite sign, because the holes and electrons are accumulated at the same side of sample, but they have opposite charge (Fig.14). Because of a nearly- equal number of holes and electrons in a metal, the Hall voltage in the metal is small and it is proportional to the gradient of the density of states at the Fermi level. When the electron gas in a metal is spin-polarized, both the electrons and holes are spin-polarized. Because of the Hall effect, the spin-polarized electrons and holes are accumulated at the same side of the sample. Therefore, at this side of sample a significant spin accumulation occurs.

The reason why in a metal the spin is accumulated due to the Hall effect can be understood as follows. As was explained in Chapter 12, in a drift current the "spin" states with energy above the Fermi energy are drifted from "+" to "-" as a negatively-charged particle should do. This current is defined as the electron current. The "spin" states with energy below the Fermi energy (see Fig. 10(a)) are drifted in the opposite direction from "-" to "+", despite they are negatively charged. This current and an additional current of negatively-charged "full" states, which is drifted from "+" to "-", are defined as the hole current. As was explained above, in a magnetic field only the direction of movement of the electrons defines the effective electrical field, which induces the Lorentz force. Therefore, the "spin" states with energies below and above of the Fermi energy experience the effective field in opposite directions. As was just mentioned, the "spin" states with energies below and above of the Fermi energy are drifted in the opposite direction with respect to the direction of an electrical field. Overall, all "spin" states turn in the same direction and they are accumulated at the same edge side of the sample. The "spin" states of the TIA assembly transport the spin. Therefore, the spin is accumulated at this edge of the sample.

The Hall effect can be described by the force term in the Boltzmann transport equations (18)

$$\left(\frac{\partial F_i}{\partial t}\right)_{force} = -\vec{F}_{Force} \cdot \nabla_p (F_i) \tag{86}$$

where

$$\vec{F}_{Force} = -e\frac{\vec{v}}{c} \times \vec{H} \tag{87}$$

In an equilibrium in the electron gas the electrons move in all direction. Since the Lorentz force only changes the electron movement direction, the Lorentz force does not change the equilibrium electron distribution $F_{i,0}$

$$\vec{F}_{Force} \cdot \nabla_p (F_{i,0}) = 0 \tag{88}$$

Eq. (88) can be verified as follows

$$\vec{F}_{Force} \cdot \nabla_p (F_{i,0}) = \left(-e\frac{\vec{v}}{c} \times \vec{H}\right)\left(\frac{\partial F_{i,0}}{\partial E} \cdot \nabla_p E\right) = -\frac{e}{c}\frac{\partial F_{i,0}}{\partial E}(\vec{v} \times \vec{H}) \cdot \vec{v} = 0 \tag{89}$$

Only if there is a drift current, the Lorentz force may change the electron distribution. In cases of the current of the running-wave electrons, substituting Eq. (30) into Eq. (86) gives

$$\left(\frac{\partial F_i}{\partial t}\right)_{force} = -\vec{F}_{Force} \cdot \nabla_p (F_{i,1}) = \left(e\frac{\vec{v}}{c} \times \vec{H}\right) \cdot \nabla_p \left(-\tau_k \cdot \vec{v} \cdot \nabla F_{i,0}\right) = -\tau_k \cdot \frac{e}{c}(\vec{v} \times \vec{H})\left(\frac{1}{m} \cdot \nabla F_{i,0}\right) \tag{90}$$

8909

Using the relaxation-time approximation Eqs. (19),(21), ignoring the term of order $H^2$ and substituting Eq (90) in Eq. (18) gives

$$-\frac{F_{i,2}}{\tau_k} - \tau_k \cdot \frac{e}{mc}(\vec{v} \times \vec{H})\nabla F_{i,0} = 0 \tag{91}$$

From Eq. (91) the Hall current can be calculated as

$$\vec{j}_{i,Hall} = \frac{e}{(2\pi\hbar)^3}\iiint \vec{v} \cdot F_{i,2} \cdot d\vec{p} = \frac{-\tau_k^2 e}{mc(2\pi\hbar)^3} \cdot \iiint \vec{v} \cdot \left[(\vec{v} \times \vec{H})\nabla F_{i,0}\right] \cdot d\vec{p} =$$

$$= \frac{-\tau_k^2 e}{mc(2\pi\hbar)^3} \cdot \iiint |\vec{v}|^2 \cdot \left[(\vec{H} \times \nabla F_{i,0})\right] \cdot d\vec{p} = \frac{-\tau_k^2 e}{mc} \cdot \int D(E) \cdot |\vec{v}|^2 \cdot \left[(\vec{H} \times \nabla F_{i,0})\right] \cdot dE \tag{92}$$

## *13. Conclusion*

The charge and spin transport in the electron gas was studied by solving the modified Boltzmann transport equations. It was shown that 5 material parameters are required to describe the spin and charge transport in a conductor. These parameters are the charge, injection, spin-diffusion and detection conductivities, the spin life time and the density of states at the Fermi energy. The conductivities can be calculated from the modified Boltzmann transport equations.

Several essential facts were included into the modified Boltzmann transport equations. It was shown that instead of one, three transport equations should be used. The transport of "full" states, "spin" states of the TIS assembly and "spin" states of the TIS assembly should be described by individual transport equations. It was shown that the electrons in the electron gas can be either of the running-wave type or the standing-wave type. The different transport mechanisms of these two types of electrons were included in the modified Boltzmann transport equations. The spin properties of electron transport depend significantly on relative amounts of each type of electrons.

The Spin Proximity effect was described. It is a well-known fact that a spin accumulation diffuses from regions of a higher spin accumulation to the regions of a smaller spin accumulation. The Spin Proximity effect describes the fact that at a contact between two conductors of different equilibrium spin polarization of the electron gas the spin accumulation diffuses from the conductor of a higher spin polarization to the conductor of a smaller spin polarization. For example, in the vicinity of a contact between a ferromagnetic and non-magnetic metals, in the ferromagnetic metal the equilibrium spin accumulation becomes smaller than in the bulk and there is a spin accumulation in the non-magnetic metal.

When a drift current flows through the contact, the spin accumulation may be drifted from one metal to the other metal. The spin accumulation becomes smaller in one metal and it becomes larger in the other metal. This effect is called the spin injection. The spin injection only modifies the distribution of the spin accumulation across the contact, which was initially established due to the Spin Proximity effect. It should be noticed that the magnitude of the Spin Proximity effect is proportional to the spin-diffusion conductivity $\sigma_{spin}$ and the magnitude of the spin injection is proportional to the injection conductivity $\sigma_{injection}$. Since in the metals $\sigma_{spin} \gg \sigma_{injection}$, often a significant drift current is required in order to modify the distribution of the spin accumulation across a contact enough so that the modification can be detected.

In a semiconductor there is only one type of carries. It is either electrons or holes. In metals both the electrons and holes contribute to the transport. The drift current in a metal consists of opposite flows of the electrons and holes. Because of this feature, there are two unique properties of the electron transport in a metal. The first property is a substantially-smaller spin-polarization of a drift current compared to the spin-polarization of the electron gas in the metal. The second property is a large spin accumulation and a small charge accumulation at one edge of a sample due to the Hall effect. The electrons and holes have opposite charge, but the same spin direction. In the drift current, the electrons and holes transport the charge in the same direction, but the spin in the opposite directions. This reason why in a metal $\sigma_{charge} \gg \sigma_{injection}$ and the spin polarization of the drift current is significantly smaller than the spin polarization of the electron gas. When a magnetic field is applied perpendicularly to the drift current, both the electrons and holes are accumulated at the same side of the sample. Since the charge of electrons and holes is opposite, the Hall voltage in metals is small. In contrary, the spin direction of the accumulated spins is the same for electrons and holes. It is the reason of the substantial enlargement of the spin accumulation at one edge of sample due to the Hall effect.

The origin of the spin detection effect is understood. A spin diffusion current without a diffusion of charge is possible in the bulk of a conductor without defects. In this case there is a balance of equal amounts of spin-polarized and spin-unpolarized electrons flowing in opposite directions. In a conductor with defects and in the vicinity of an interface this balance is broken and there is a charge diffusion along the spin diffusion. This causes a charge accumulation along the spin diffusion. The voltage induced by this charge accumulation can be measured and the magnitude of the spin current and spin accumulation can be evaluated.

The conventional charge conductivity and spin-diffusion, detection and injection conductivities are not fixed and unchangeable material parameters. They are changing significantly depending on the density and the distribution of spatial defects in a conductor. They are different in the bulk of a conductor and in the vicinity of an interface. For example, the detection conductivity is zero in the bulk of a metal or a semiconductor, but it can be substantial in the vicinity of a contact. Even the transport mechanism may change in the vicinity of a contact. If the major transport mechanism in a metal is the current of the running-wave electrons, the major transport mechanism becomes the scattering current, for example, in the vicinity of a tunnel junction.

# References

*Electronic address: v.zayets@aist.go.jp. Web: http://staff.aist.go.jp/v.zayets/
[1] C. Kittel, Introduction to Solid State Physics, Wiley: New York, 1996;
[2] C. Zener, Phys. Rev. 81, 440 (1950);
[3] E.C. Stoner, Proc. Roy. Soc., **A165**, 656 (1938);
[4] V. Zayets *J. Mag. and Mag. Mat.* **356** 52-67 (2014) 52–67;
[5] T. Valet and A. Fert, Phys. Rev.B. 48, 7099 (1993);
[6] V.Zayets, Phys. Rev.B. **86**, 174415 (2012);
[7] M.J. Zuckermann, Solid State Commun., **12,** 745 (1973);
[8] P. K. Manna, S. M. Yusuf, Phys. Report-Rev. **535**, 61-99 (2014);
[7] M.J. Zuckermann, Solid State Commun., **12,** 745 (1973);
[8] P. K. Manna, S. M. Yusuf, Phys. Report-Rev. **535**, 61-99 (2014);
[9] M.M. Schwickert, R. Coehoorn, M.A. Tomaz, E. Mayo, D. Lederman, W.L. O'Brien, T. Lin, G.R. Harp, Phys. Rev. B **57** (1998) 13681.
[10]. S. A. Crooker, M. Furis, X. Lou, C. Adelmann, D. L. Smith, C. J. Palmstrom, and P. A. Crowell, Science **309**, 2191 (2005).
[11]. M. Furis, D. L. Smith, S. Kos, E. S. Garlid, K. S. M. Reddy, C. J. Palmstrom, P. A. Crowell, and S. A. Crooker, New Journal of Physics **9**, 347 (2007).
[12] T. Suzuki, T. Sasaki1, T.Oikawa, M. Shiraishi, Y. Suzuki, and K. Noguchi, Applied Physics Express **4**, 023003 (2011).
[13] F. J. Jedema, A. T. Filip, and B. J. van Wees, Nature **410**, 345 (2001).
[14] M.I. Dyakonov and V.I. Perel, Phys. Lett. A **35**, 459 (1971).
[15] Y. Kato, R. C. Myers, A. C. Gossard, and D. D. Awschalom, Science 306, 1910 (2004).
[16] Y. Ohno, D. K. Young, B. Beschoten, F. Matsukura, H. Ohno, and D. D. Awschalom, Nature **402**, 790 (1999).
[17] R. Fiederling, M. Keim, G. Reuscher, W. Ossau, G. Schmidt, A. Waag, and L. W. Molenkamp, Nature **402**, 787 (1999).
[18] V. F. Motsnyi, V. I. Safarov, J. De Boeck, J. Das, W. Van Roy, E. Goovaerts, and G. Borghs, Appl. Phys. Lett. **81**, 265 (2002).
[19] B.T. Jonker, G. Kioseoglou, A.T. Hanbicki, C.H. Li and P.E. Thompson, Nature Physics **3**, 542 (2007).
[20] E. Yablonovitch (1987), Phys. Rev. Lett. **58,** 2059–2062 (1987).